# Beam shaping techniques for pulsed laser ablation in liquids: Unlocking tunable control of nanoparticle synthesis in liquids

S. Molina-Prados[1], N. M. Bulgakova[2], A. V. Bulgakov[2], J. Lancis[1], G. Mínguez Vega*[1], C. Doñate-Buendia*[1]

[1]GROC-UJI, Institute of New Imaging Technologies, Universitat Jaume I, Castellon, Spain.

[2]FZU - Institute of Physics of the Czech Academy of Sciences, Prague, Czech Republic.

*Corresponding authors: Carlos Doñate Buendia – cdonate@uji.es and Gladys Mínguez Vega – gminguez@uji.es

## Abstract

Nanoparticle synthesis via pulsed laser ablation in liquids has gained prominence as a versatile and environmentally friendly approach for producing ligand-free colloids with controlled composition, size, and morphology. While pulsed laser ablation in liquids offers unparalleled advantages in terms of nanoparticle purity and material versatility, enhancing the size control and productivity require modifications of the standard pulsed laser ablation in liquid technique such as the incorporation of beam shaping techniques. Recent developments in spatial and temporal beam shaping have demonstrated their potential to revolutionize pulsed laser ablation in liquids by enabling more precise energy deposition and modified nanoparticle production dynamics. This review highlights the critical role of beam shaping, encompassing spatial shaping of the beam to influence laser-material interaction, and temporal modification to optimize pulse duration and energy delivery. The current advancements in beam shaping techniques, their impact on the nanoparticle characteristics, and their broader implications for scaling pulsed laser ablation in liquids to meet industrial demands are highlighted, offering a comprehensive perspective on the future of this dynamic field.

## 1. Introduction

Laser ablation in liquids (PLAL) [1–5] is an increasingly employed nanoparticle synthesis technique first established in the 1990s [6,7]. This method involves focusing high-energy laser pulses onto a solid target submerged in a liquid medium [8]. As the laser interacts with the target, it triggers rapid ionization, heating, and evaporation of the material, leading to plasma formation. The plasma cools down in the surrounding liquid releasing some nanoparticles (NPs) into the liquid, the cooling process also generates gas bubbles from the liquid environment. These gas bubbles nucleate, forming a cavitation bubble (CB). Additional NPs are formed within this bubble until it collapses, releasing and ejecting the remaining NPs into the liquid [9–11]. Although nanomaterials can be produced by alternative physical, chemical or biological methods [12], PLAL offers several advantages [13]. Chemical methods are effective at controlling NP size and shape but often require reducing agents and stabilising or capping agents to ensure colloidal stability, which may introduce impurities and raise environmental concerns. Biological synthesis routes for synthesizing nanomaterials, such as plant extracts, bacteria, or fungi-based ones, are indeed eco-friendly and offer advantages like low toxicity and reduced environmental impact but generally offer less control over NP properties. Moreover, other physical methods such as milling, pyrolysis, sputtering, arc discharge or gas-phase processes such as flame spray pyrolysis or gas aggregation are highly scalable and productive, though they typically require high temperatures or vacuum systems and may offer limited control over surface chemistry.

Compared with these approaches, PLAL provides high-purity colloids with reduced impurities and byproducts, even allowing tuning of NP size, crystallinity, defects, and optical properties. Conventional batch PLAL setups often exhibit lower productivity than large-scale chemical or gas-phase synthesis; however, recent advances using MHz-repetition-rate lasers and continuous flow configurations have significantly increased yield, demonstrating that scale-up is feasible and cost-effective [14].

Figure 1 provides a comparative overview of these methods considering key factors: elemental flexibility, environmental friendliness, synthesis purity, health and safety, productivity, process complexity, and degree of remote controllability. The latter refers to the ability to fully automate and operate the process under closed-loop control. PLAL lends itself naturally to remote monitoring and automation through computer-controlled lasers, scanning systems, and online spectroscopy, [15] although similar automation can also be achieved in chemical synthesis using microfluidics-based reactors [14].

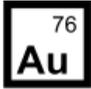

**Figure 1.** Comparative analysis of the main features of nanomaterial synthesis techniques, i.e. chemical, biological, physical and laser synthesis.

PLAL's mechanisms and the ablation dynamics can be described starting with the pulse emission from the laser source [16]. The laser beam travels through the transparent liquid layer, ideally minimizing energy losses due to absorption and avoiding non-linear optical effects [17]. Upon reaching the target, the laser pulse induces rapid electronic excitation, leading to the injection of electrons into the surrounding liquid, within the first few picoseconds, from 1 to 20 ps [18]. This triggers the formation of a dense plasma composed of the target material components, which remains active for 20 to 200 picoseconds [19,20]. The plasma expands rapidly, generating a mechanical shockwave in both the target and the liquid, with pressures at the laser's focal spot reaching tens of gigapascals, but strongly depending on the specific laser conditions [21]. The high pressures induce spallation of the target surface, while the plasma interacts with the liquid, vaporizing part of it and forming a cavitation bubble (CB) on a nanosecond timescale [22–25]. Throughout microseconds, the CB grows and collapses, releasing NPs into the liquid environment [9], Figure 2a.

The early stages of laser ablation, along with variations in material density, temperature, and phase states, can be effectively modelled using large-scale atomistic simulations [26,27].

PLAL is a simple, fast, and versatile technique that has been employed to produce ligand-free NPs [28], core-shell structures [29,30], heterostructures [31], nanoalloys [32,33], hybrid materials [34,35], and complex multi-element nanomaterials such as high-entropy alloys [36–38]. This method enables the synthesis of nanomaterials with exceptional purity from virtually any solid target [39], including pure metals [40–43], semiconductors [44–47], or dielectric materials [48,49], in organic [41,50] or inorganic solvents [41,50–52]. Although initial NP production rates via PLAL were limited to a few milligrams per hour, recent advancements in high-power, high-repetition-rate laser systems [53,54], as well as strategies such as the use of diffractive optical elements to create multifocal structures [55,56], CB bypassing [57], reducing the liquid layer [58,59], avoiding non-linear effects [60], and optimizing sample geometry [61], have significantly increased the production rates up to several grams per hour in specialized setups requiring fast scanning systems and high power (500 W) and repetition rate (10 MHz) picosecond laser sources [5,35,53,54]. Only recently, the employment of diffractive optical elements has allowed to reach the gram per hour productivity scale with industrially available picosecond laser sources, still requiring high power sources (100 W) but removing the requirement for high repetition rate and faster scanning speeds [55].

PLAL-produced nanomaterials have broad applications across different nanotechnology fields, including X-ray radiotherapy [62], boron neutron capture therapy [62,63], viral [64,65] and microbial growth inhibition [42,66], antibacterial agents [67,68], anticancer treatments [67,69], magnetic resonance imaging contrast agent [70], photothermal therapy [71,72], cell imaging [73], proton therapy enhancement [74,75], fluorescence [76,77] and colorimetric sensors [78,79], surface-enhanced Raman spectroscopy detection [71,80,81], nanofluids for thermal applications [82–84], additive manufacturing [85–87] or catalysis [88,89]. The previously mentioned applications of PLAL-derived NPs can be grouped into four major categories: catalysis [90], advanced materials [91], sensing and filtration [92], and bio-applications [93]. Figure 2b schematically illustrates this classification.

However, several limitations remain that hinder its widespread industrial adoption, including the overall yield and scalability, particularly when attempting to synthesize complex nanostructures at industrial volumes, as well as the reproducibility and precise control over NP properties such as shape. As discussed by Jendrzej et al. [14], when scaling up PLAL, the cost of the laser becomes a negligible factor compared to process efficiency, throughput, and colloid stability. Therefore, optimising ablation conditions (e.g., using high-repetition-rate lasers, flow-through cells, and improved cavitation management) will be critical for enabling commercially viable production [94].

One promising strategy to overcome these limitations involves controlling the spatial and temporal profiles of the laser beam. Spatial beam shaping allows for fine adjustment of energy distribution on the target, influencing ablation efficiency and NP uniformity. Figure 3 provides an overview of representative approaches and the corresponding beam profiles they enable. Similarly, temporal pulse shaping modifies the interaction time and heat generation and dissipation, allowing for better control of overheat accumulation and nonlinear effects during ablation. These approaches provide a path toward more consistent and tunable NP synthesis, further unlocking the potential of PLAL.

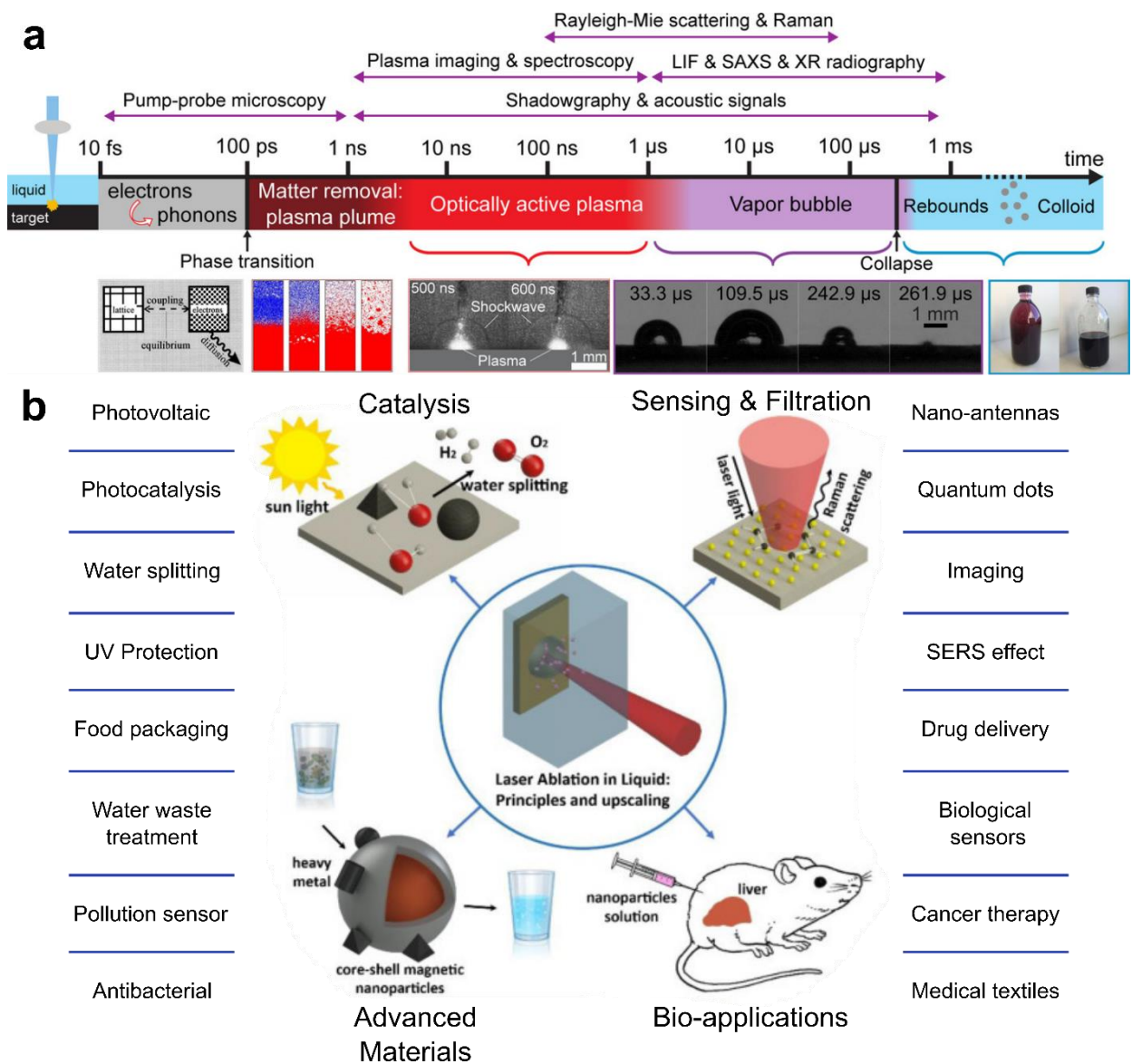

**Figure 2.** a) PLAL processes timeline starting from the laser pulse interaction with the target until release of the NPs and colloid formation. On top, the temporal and spatially resolved characterization techniques are displayed according to their temporal resolution. The figure is reprinted from [29] Copyright 2020, under Creative Commons Attribution. Retrieved from https://doi.org/10.1002/chem.202000686. b) PLAL NPs applications, arrows point outwards to the four defined areas, catalysis, advanced materials, sensing and filtration, and bio-applications. The figure is reprinted from [2] Copyright 2020, under Creative Commons Attribution. Retrieved from https://doi.org/10.3390/nano10112317.

# 2. Spatial and temporal beam shaping in material processing

The spatial focusing of the laser beam on the sample directly influences the ablation efficiency and the quality of the produced NPs. The most widely used PLAL system in different laboratories around the world consists on a convergent lens with the laser beam perpendicular to the target [5]. However, the laser ablation process is highly sensitive to both the spatial and temporal profiles of the beam. While there are already some studies exploring the effects of beam shaping, this field remains largely underexplored. Further advancements in spatiotemporal beam shaping techniques hold significant potential to greatly enhance control over NP synthesis by PLAL.

In the context of PLAL, the NP yield is typically defined as the mass of NPs collected per unit of time (mg·h$^{-1}$) or per unit of laser energy (mg·J$^{-1}$). The ablation rate corresponds to the total mass removed from the target per unit of time or per pulse (mg·h$^{-1}$, mg·pulse$^{-1}$), while the collection efficiency denotes the fraction of this ablated mass ultimately recovered as colloidal NPs. Quantification can be carried out by gravimetric analysis of the colloid, either collecting the entire suspension, evaporating the solvent to dry it, and weighing the residue, or by measuring target mass loss before and after ablation, which allows estimation of collection efficiency. Greater precision is obtained by Inductively Coupled Plasma-based (ICP) techniques (ICP-MS, ICP-OES, AAS), particularly for metallic targets, as they determine the concentration in a representative colloidal aliquot. Complementary methods include UV–Vis spectroscopy with calibration standards [95] or thermogravimetric analysis (TGA) in the presence of organic species or adsorbates [73].

When reporting NP yields, several practical issues must be considered. Collection efficiencies are generally below unity, meaning that the ablated mass does not directly equal the NP mass in suspension. Incomplete solvent evaporation or residual organics can bias gravimetric results, making vacuum drying preferable. NP adhesion to container walls requires vessel washing, and aggregation during or after ablation may need sonication or size-separation steps. Overall, gravimetry provides accessible but sometimes overestimated values, whereas ICP-based approaches offer more accurate and reproducible quantification, facilitating reliable comparison of different PLAL conditions.

From a spatial perspective, the beam profile can be modified from the standard Gaussian profile into top-hat, doughnut-shaped, or multiple-beams using optical elements, including cutting-edge technology such as metamaterials, or refractive, reflective, or diffractive elements [96,97]. Beam shaping can be accomplished through a variety of advanced optical technologies, applied either individually or in combination, as shown in Figure 3. As a general strategy for imaging and material processing, beam shaping technologies represent a research field covered in numerous reviews [98,99]. In subsequent sections, this review offers a more detailed discussion of beam-shaping methodologies for nanoparticle synthesis and the underlying optical mechanisms including its influence on ablation efficiency, size control, and colloidal stability.

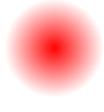

**Figure 3.** Schematic illustration of spatial beam shaping strategies in PLAL. Different beam profiles can be generated from the standard Gaussian mode: top-hat, doughnut-shaped, and multiple-beam configurations. Such modifications influence the local fluence distribution on the target, thereby affecting ablation efficiency, NP size distribution, and colloid uniformity.

Refractive elements rely on the interaction of light with a transparent material with a different refractive index than air. Standard refractive optics elements such as lenses often suffer from aberrations and can introduce temporal pulse broadening due to the wavelength dependence of the refractive index. Alternative refractive optics technologies include dynamic lenses and freeform optics (both in refraction and reflection) [100]. Unlike traditional lenses, freeform optics can generate complex and exotic light structures that go beyond simple focusing, allowing the design of custom-shaped intensity distributions at the target. This technology permits the creation of arbitrary beam shapes, enabling more intricate control of the laser interaction with the material, which is especially useful for advanced applications requiring customized energy deposition patterns. Diffractive elements offer the advantage of being thin and lightweight and allow the generation of user-defined beam patterns through controlled light diffraction produced by micro or nanostructured glasses or surfaces. However, they are specifically designed for a single wavelength, and their use with ultrashort pulses can lead to substantial temporal pulse broadening [101].

Laser beam shaping technologies can be also categorized into static and dynamic systems. Static systems include components such as microlens arrays, cylindrical lenses, diffractive lenses, flat optics, etc. which provide fixed spatial beam profiles [102]. For dynamic beam control, spatial light modulators (SLMs) are employed, with liquid crystal modulators, membrane mirrors, and digital micromirror devices (DMDs) being the most employed technologies [103]. Liquid crystal modulators are well-suited for modulating both amplitude and phase [104], but they operate at relatively slow speeds, typically in the range of a few hertz (Hz). In contrast, DMDs offer very high modulation speed, though they encode information in a binary format, limiting their flexibility for some applications. Membrane mirrors, on the other hand, allow for complete control of the beam, i.e. intensity and full surface actuation. Nevertheless, their low number of actuators restricts the precision of the modulation, making them primarily useful for correcting optical aberrations. Although SLMs have demonstrated their relevance in material processing, their application in PLAL remains largely underdeveloped and presents significant opportunities for further exploration.

From the temporal pulse perspective, the main strategies for material processing are the generation of double pulses or pulse bursts [105], as well as modifying the temporal profile of the pulsed beam. In the first case, the laser technological developments in the last decade have allowed the generation of megahertz and gigahertz bursts using acousto-optic devices within the laser cavity. In the second case, pulse compressors enable the creation of user-defined temporal pulse profiles in Fourier space shaping [106] or direct space-to-time pulse shapers [107,108]. Temporal pulse shaping has been employed in laser material processing to reduce the required laser fluence for processing and to maximize ablation efficiency through minimization of thermal dissipation [109]. However, pulse shaping in PLAL still represents a barely explored approach, being mostly limited to experiments with controlled temporal delays between pulses. Further evaluating the possibilities of pulse shaping in PLAL holds the potential to provide enhanced control over laser-matter interactions, facilitating more precise NP synthesis.

# 3. Advanced Spatial Beam Shaping in PLAL

## 3.1. The role of spatial beam shaping in PLAL

The spatial profile of the laser beam plays a pivotal role in influencing all stages of the PLAL process. It governs critical features of PLAL, from the laser interaction with the liquid and the target to the subsequent modifications in the shape and dynamics of the CB, ablation rate, depth and area. These interdependent processes ultimately define the properties of the NPs, such as their size and distribution. The influence of these factors and their interrelations will be discussed in this section emphasizing the role of the irradiance, also called intensity, (W/cm$^2$) and the fluence (J/cm$^2$) of the laser beam. Both variables strongly depend on the beam size and shape and fundamentally describe the effect of beam shaping compared to a parameter like pulse energy that remains constant if the spatial beam profile is modified.

When the laser enters the liquid, it can be absorbed, i.e. water and most of the organic solvents absorb radiation for wavelengths in the near infrared, this produces heat that can cause a vaporization of the liquid layer once the fluence threshold of vaporization is reached [59]. At high irradiances, in the $GW/cm^2$ range, nonlinear optical effects such as self-focusing, supercontinuum generation, multiphoton ionization, filamentation or optical breakdown can become significant [59,60,110]. These effects can influence how the beam propagates through the liquid and interacts with the target and have a strong dependence on the beam profile [111]. Self-focusing is a non-linear optical process induced by a local modification of the refractive index of a material due to the propagation of an intense laser beam due to the Kerr-effect [59,112]. Self-focusing occurs in water medium at an irradiance threshold of $0.5 - 1 \times 10^{10} W/cm^2$ [113]. When the irradiance increases exceeding the self-focusing threshold, beyond $10^{10} W/cm^2$, several nonlinear processes such as self-phase modulation, four-wave mixing, Raman scattering, and self-steepening, cause severe spectral broadening, generating a supercontinuum. In addition to self-focusing on the solvent it is important to consider the influence of NPs already present in the liquid on the optical breakdown threshold. Dispersed NPs act as additional scattering and absorption centres, leading to local field enhancement in their vicinity. These effects facilitate multiphoton absorption and avalanche ionization processes, effectively lowering the breakdown threshold of the liquid compared to NP-free conditions. This is shown by the fluence dependence of the generation rate of hydroxyl radicals during optical breakdown in water with the presence of on terbium NPs. When the fluence increases from ~60 to 140 J·cm⁻², the radical yield with oxidised NPs rises by over an order of magnitude [114,115]. Moreover, according to Davletshin et al. plasmonic coupling in aqueous Au nanospheres leads to enhanced local fields. This leads to a reduction via near-field amplification of up to four orders of magnitudes in the effective optical breakdown threshold under 3 ps irradiation, compared to the base liquid without NPs ($8.5 \times 10^{11} W/cm^2$) [116,117].Several strategies both statical [118] and dynamical [119] have been considered to modify the spatial beam profile of the laser with an impact on the supercontinuum effect. Considering that the supercontinuum can modify the size of the NPs in a colloid due to fragmentation [120–122], it can be a tool to control NP distribution during or after PLAL.

For irradiances that exceed 10¹² W/cm², laser filamentation can occur. It is a complex process that results from the dynamic balance between self-focusing and plasma-defocusing. An example of filamentation produced in ethanol in a PLAL setup can be observed in Figure 4 [123]. Initially, high-intensity laser irradiation inside the liquid medium triggers' multiphoton ionization and tunnel ionization, generating free electrons due to the high peak power. The excess of electrons and ions induced by the multiphoton ionization [121] have been used in reactive ablation for the fabrication of NPs [124,125] and hybrid colloidal NPs [126]. In other applications, dynamic control of the filaments can be achieved by employing spatial light modulation [127].

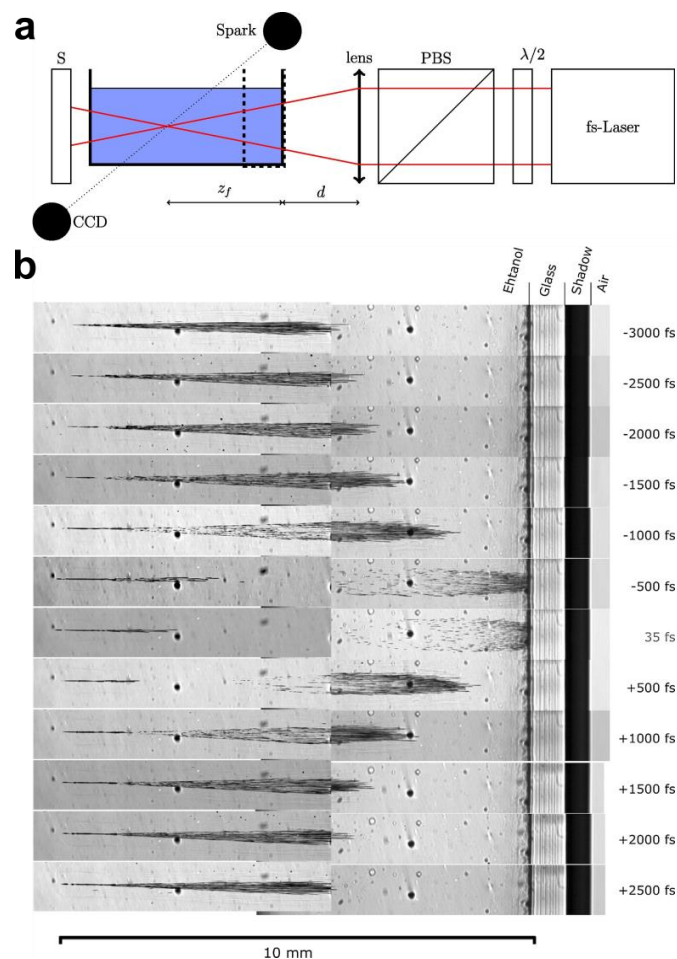

**Figure 4.** a) Experimental setup for the record of filamentation produced in a liquid. b) Laser focusing on ethanol, the laser is on the right-hand side. Pulse energy $E_p = 1 mJ$, wavelength λ=800 nm. Several images corresponding to different pulse durations are shown one above the other. Reprinted from [123] Copyright (2018), with permission from Elsevier.

Finally, the optical breakdown irradiance threshold in aqueous media is 1.11×10$^{13}$ W/cm$^2$ [128]. This process is physically observed by bubble formation in the liquid [129]. The dissociation of the liquid due to multiphoton ionization leads to the formation of hot plasma with temperatures reaching 10$^4$ K. Subsequently, plasma recombination begins, and the high-temperature plasma is replaced by vaporized fluid, leading to the creation of microbubbles and mechanical effects such as shock wave emission and cavitation. In general, these microbubbles cause laser energy losses and distortion of the laser spatial profile due to scattering and should be avoided to maximize PLAL productivity and control the irradiation parameters on the target.

The fluence ($\Phi_0$), energy per unit area, is another key parameter of the laser that can be modified by spatial shaping. The fluence directly influences most of the properties of the ablation process as the size of the CB, where the maximum bubble volume scales linearly with the laser fluence for the high-fluence range, $100 - 200 \, J/cm^2$ [130]. The NP size distribution is also affected by the fluence, obtaining smaller NPs for low fluences, $0.5 - 4 \, J/cm^2$ [131–133]. The NP composition is also affected by the fluence, an increase of the laser energy leads to a higher concentration of gas, for example oxygen atoms, that increase O content in the produced NPs [134] and affecting the morphology [135,136]. Regarding the ablation efficiency, a maximum is reported for ablation in air for $\phi_0 = e^2 \phi_{th}$, being $\phi_{th}$ the ablation threshold [53]. In liquids, ablation efficiency is usually determined experimentally due to the energy losses and beam modification due to confinement, plasma–liquid interactions, and the cavitation bubbles. Dittrich *et al.* found that the ablation threshold for Au in air is ~1.9× higher than in water [137], and Sun *et al.* reported a reduction from ~2.22 J/cm$^{-2}$ (air) to ~1.02 J/cm$^{-2}$ (liquid-assisted) [138]. Therefore, the air-based formula is often retained but complemented by an empirical correction factor to account for liquid effects as proved by Intartaglia et al. for Si NPs production with a picosecond laser [139] and Kanitz et al. for Fe NPs production with femtosecond laser [20]. Up to now, control of the fluence has been performed by varying the laser pulse energy [133,140]. In the last years, other approaches to modify the fluence based on the change in the spatial profile have emerged [135,141–143].

## 3.2. Focal spot shape conformation in PLAL

The interaction of the laser with the target and the generated colloid depends not only on the beam fluence but also on the shape. The focusing systems employed in the synthesis of the NPs affect their properties. Therefore, this section will review the main beam-shaping techniques that are being considered in PLAL for NP formation influencing. We will divide the section into two different subsections. In the first one, we will consider spatial beam profile variations in conventional PLAL systems (defocusing, tilting and the incorporation of cylindrical lenses). Meanwhile, in the second subsection, we will focus on systems where an external element is added to modify the beam profile into a non-gaussian beam (bessel beam, doughnut and speckle pattern).

### 3.2.1. Conventional Spatial Beam Control

Although the determination of the focused laser beams in air can be done directly from the ablated spots at different lens-target distances, in liquids and with ultrashort laser pulses this procedure is not so evident. The diameter of the ablated spots under different water layers for a 120 fs pulsed beam with a spherical lens of 40 mm focal length is shown in Figure 5. When ablation is carried out in air, the craters caused in the target have the smallest diameter when the relative geometric focal length of the lens coincides with the target, $d_{z_L} = 0$. The diameter of the craters increases when the relative position of the lens-objective system moves towards the target ($d_{z_L} < 0$) or away from it ($d_{z_L} > 0$). Consequently, the smaller the ablated spot, the higher the fluence, and a higher productivity can be achieved [59]. In comparison, when ablation takes place within a liquid, the ablated spot diminishes until vanishing with the increment of the lens position $d_{z_L}$. However, when NPs are generated in liquid using femtosecond pulses, the largest ablated mass does not occur for the lens-target position corresponding to the smallest ablated spot. In fact, the maximum NP productivity with a 10 mm liquid layer was achieved for $d_{z_L} = +2\ mm$. According to Figure 5, the smallest spot found for these parameters was at $d_{z_L} = +4\ mm$. This 2 mm difference in the lens position induces a 260 times higher Au colloid concentration. For this reason, defocusing is a simple, alternative technique to increase NP production by modifying the energy distribution when ablating the target. Defocusing plays a dual role depending on whether the local fluence is maintained. When the beam is defocused without power compensation, the fluence at the target decreases, leading to lower plasma temperatures and reduced ablation efficiency, which generally results in smaller NPs with narrower size distributions.

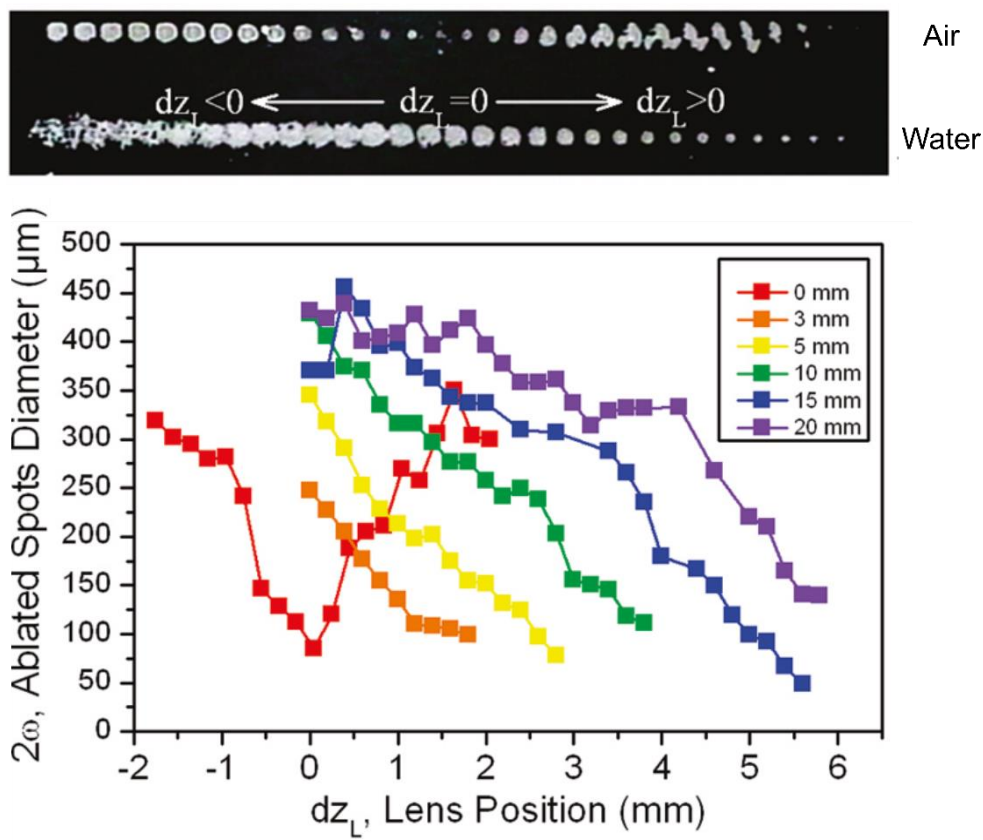

**Figure 5.** Top: Ablated spot diameter on a silicon wafer comparison between air and with 10 mm water layer. Bottom: Effect of lens position using a spherical lens with 40 mm of focal length in different liquid layers. Reprinted with permission from [59]. Copyright 2011, American Chemical Society.

When the laser power is increased to compensate for the defocusing the fluence is kept constant, but a larger ablation area is irradiated, which increases NP productivity. The higher colloidal concentration enhances aggregation, often resulting in larger mean particle sizes. [144]. Furthermore, the precision of the focusing can play a crucial role in the physicochemical properties of the colloids. Ryabchikov et al. have demonstrated that the optical properties of Si/Au NPs, their structure as well as their chemical composition can be modified by defocusing. Defocusing 0.5 mm inside the target led to enhanced chemical stability of the colloids and increased concentration. Moreover, NP size control could be achieved by defocusing. The hydrodynamic diameter increases with defocusing, with the smallest diameter achieved when the focal spot is placed on the target surface, Figure 6. Therefore, the reported increase or decrease in NP size under defocusing is not contradictory but rather reflects the different strategies used to control fluence during the process.

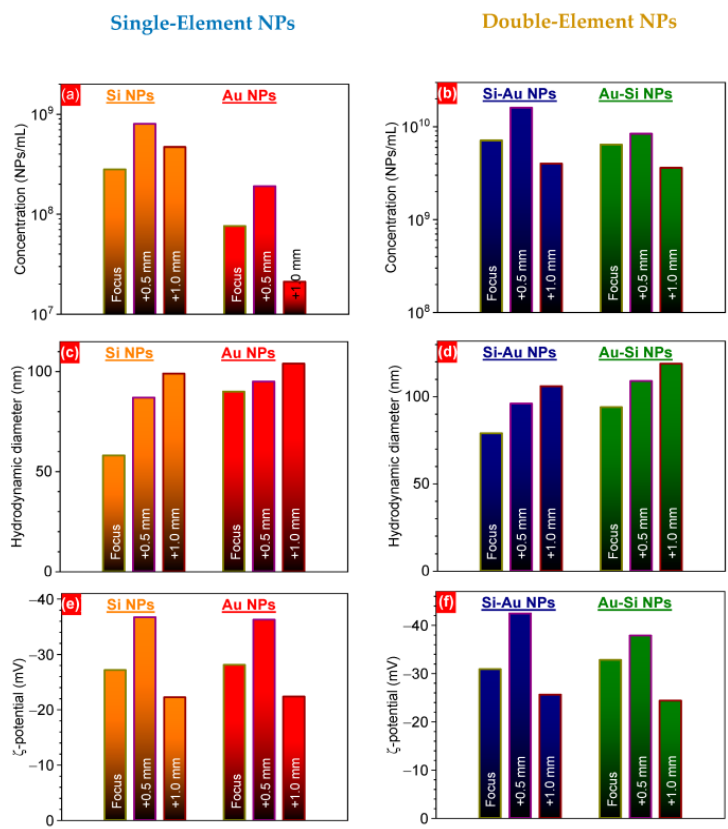

**Figure 6.** Nanoparticle concentrations (**a**,**b**) hydrodynamic diameters (**c**,**d**) and ξ-potential of one and two-component NPs prepared by PLAL, at different focus positions. The figure is reprinted from [144] Copyright 2025, under Creative Commons Attribution. Retrieved from https://doi.org/10.3390/cryst15020132.

Another way to modify the beam shape without changing the PLAL system is either by tilting the sample or the incident beam. The effect of target tilt along the laser irradiation direction has been investigated by Al-Mamun et al by producing spherical $Al_2O_3$ NPs by nanosecond laser ablation in water [142]. Tilting the target results in smaller particles with a narrower distribution due to the larger spot and lower fluence [142]. The morphology and shape of the ablated NPs remain constant. The produced NPs were spherical with an average particle size ranging from 8 to 18 nm for different laser parameters.

Conventionally, in PLAL spherical lenses are used. Unlike ablation with spherical lenses, ablation with cylindrical lenses modifies the spot size into an elliptical shape, increasing the area. Podagtlapalli et al. demonstrated that the use of cylindrical lenses in PLAL can lead to the formation of nanoribbons [141], and it may be attributed to two main mechanisms. First, the Ag nanospheres initially formed could undergo nano-welding under the influence of a line-shaped light sheet produced by cylindrical focusing, leading to chain- or ribbon-like structures. Second, the cavitation bubble generated during ablation and its subsequent oscillations play a critical role. Under cylindrical focusing, the cavitation bubble adopts an elongated shape that differs from the typical spherical morphology observed with circular beams [145]. This anisotropic geometry alters the internal pressure distribution and collapse dynamics, affecting the interaction between the plasma plume, shock waves, and the surrounding liquid. Consequently, both the light-induced welding and the anisotropic cavitation bubble dynamics are key factors determining the formation of the elongated nanostructures observed under cylindrical focusing.

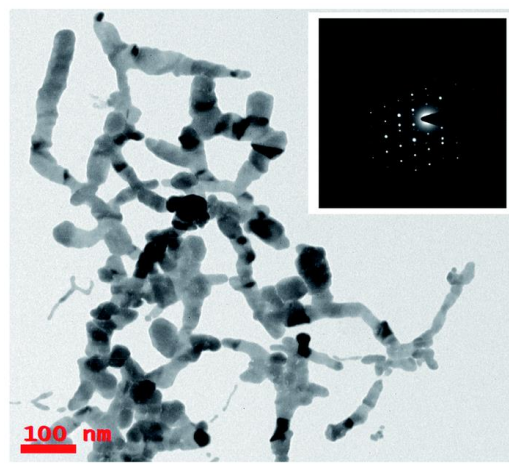

**Figure 7.** TEM images of Ag nanoribbons produced by cylindrical lens ablation at a pulse energy of 1,2 mJ. Ag nanowires/ribbons with an average length of 0.5–1 μm. The inset shows the SAED patterns, which reveal the crystalline phase of Ag nanowires. The figure is reprinted from [141] Copyright 2020, under Creative Commons Attribution. Retrieved from https://doi.org/10.1039/D0RA05942K.

The cylindrical lens PLAL experiments were performed with a 2 ps laser and a cylindrical lens of 45 mm focal length, using pulse energies between sensing and filtration 1.0 and 1.4 mJ at 800 nm wavelength. It was observed that at 1.2 mJ pulse energy, the fabrication of Ag nanoribbons longer than 0.6 μm were synthesized Figure 7.

**Defocusing**
- Simple, low-cost method
- Can enhance productivity (×260 Au concentration at $d(z_L) = +2\ mm$ with 10 mm water layer)
- Allows tuning NP size (smaller at lower fluence, larger when fluence is maintained).
- Requires careful optimization of focus position and can difficult repeatability.

**Target tilt / beam tilt**
- Produces smaller, more uniform NPs.
- Easy to implement without additional optics.
- Reduced fluence on target.
- Limited impact on NP morphology (mostly spherical).

**Cylindrical lenses**
- Convert spot into an ellipse/line.
- Enable anisotropic nanostructures (Ag nanoribbons 0.5–1 μm).
- Exploit cavitation bubble oscillations.
- Restricted applicability for isotropic NPs.
- Sensitive to alignment and focusing conditions.

### 3.2.2. External Elements for Spatial Beam Shaping

The beam spatial profile plays a crucial role in laser-matter interaction processes, influencing the efficiency, precision, and morphology of the ablated material. Therefore, introducing external optical elements that modify the laser beam profile represents an interesting approach to provide further control to the PLAL process.

One of the beam profiles generated by external optical elements are Bessel beams. Bessel beams are widely used in various optical and engineering applications due to their ability to propagate long distances almost without diffraction. Compared to Gaussian pulses, the Bessel beams have a much longer focus and can be considered *optical needles*. Quasi-Bessel zero-order beams can be created by axicon optics (conical lenses), which generate interference of an incoming Gaussian laser beam along the optical axis, thus forming a long focal line. The radial distribution of intensity in a zeroth-order Bessel beam represents a central bright spot surrounded by rings of much smaller intensities so that their spacing can be approximated by the Bessel function $J_0$ [146]. The anatomy of the quasi-Bessel beam propagation in space is shown in Figure 8 with a detailed view of its spatiotemporal shape including the evolution of the cross-sectional structure [147]. Two laser beams, CW and a femtosecond laser, have been compared showing very similar structures, underlying the long effective working distances of the intense laser *needle*.

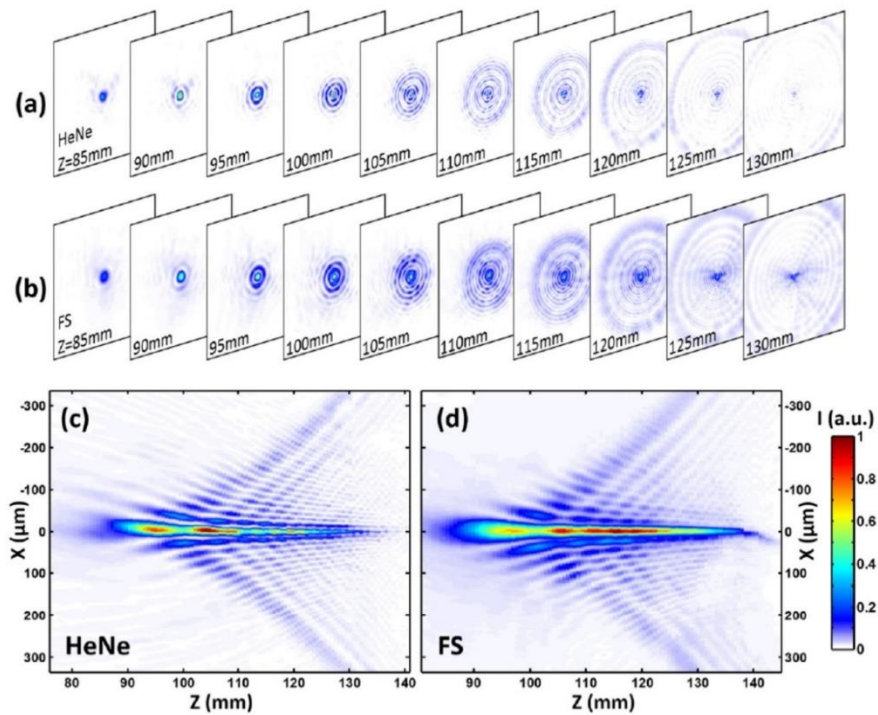

**Figure 8.** a) and b) Transverse distributions of Bessel-like beams observed at various distances (*Z*) from the axicon for a) CW HeNe and b) femtosecond (795 nm wavelength, 35 fs pulse duration) lasers. c) and d) Longitudinal profiles of CW and femtosecond laser beams respectively. Intensities are normalized to the peak values in the two cases. The figure is reprinted from [147] Copyright 2017, under Creative Commons Attribution. Retrieved from https://doi.org/10.1364/OE.25.001646.

For PLAL, the Bessel beams are still not widely used. This can relate to a very small effective diameter of the beam on the material surface. However, an axicon-generated laser beam has been shown to allow better control of NP size and offer an advantage in terms of alignment and reproducibility due to the long length of the central lobe of the Bessel beam [143]. A narrow size distribution of Ag NPs obtained by fs Bessel beam irradiation of a silver target in acetone is shown in Figure 9. As an example, the Bessel-beam-processed surface results in a giant enhancement of the SERS signal of the explosive molecule CL-20 [143]. Thus, NP synthesis in liquids using Bessel beams can provide an additional opportunity for control over colloidal properties.

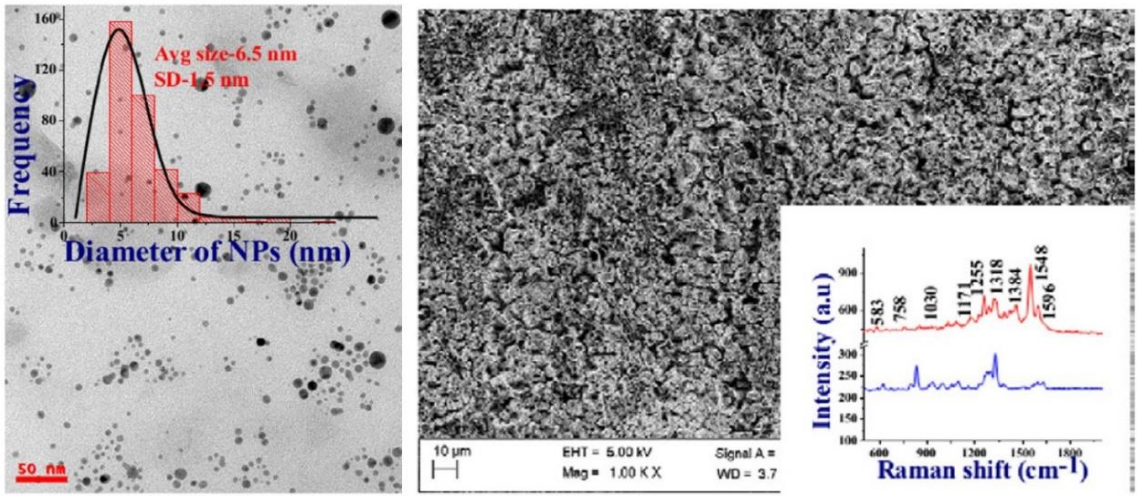

**Figure 9.** Left: TEM image of Ag colloid prepared by laser ablation in acetone by femtosecond Bessel beams at pulse energy of 1000 µJ. Right: Field emission scanning electron microscope image of Ag target processed with Bessel beam at energy 1000 µJ. Inset shows the SERS spectrum of explosive molecule CL-20 at 5 µM concentration (red curve) from the laser-exposed Ag surface. For comparison, the blue curve represents the 0.1 M CL-20 Raman spectra obtained from silicon wafer. Used with permission of Astro Ltd., from[143]; permission conveyed through Copyright Clearance Center, Inc.

The doughnut-shaped (DS) ring-like beams (also called dark-hollow beams) represent a wide class of spatially shaped light beams with controlled ring intensity and zero central intensity, i.e., with phase singularity at the beam center. Depending on the polarization, the DS beams can be classified into two sub-classes, those that carry optical angular momentum (OAM) and those that do not. The latter ones are usually radially or azimuthally polarized and, due to the cylindrical symmetry polarization, they are also called cylindrical vector beams. The beam carrying OAM has typically circular polarization and represents a sub-beam class of vortex beams. For instance, radially polarized DS beams can be focused on a spot size significantly smaller than conventional Gaussian beams [148,149] and avoid overheating of the central irradiated regions [150,151].

DS beams appear particularly attractive for PLAL with ultrashort laser pulses due to the high damage threshold and the possibility of simply switching from radial to azimuthal polarization. To generate vortex beams with their helical/twisted wavefronts, spiral phase plates are typically used to produce beams with pre-defined OAM mode contents.

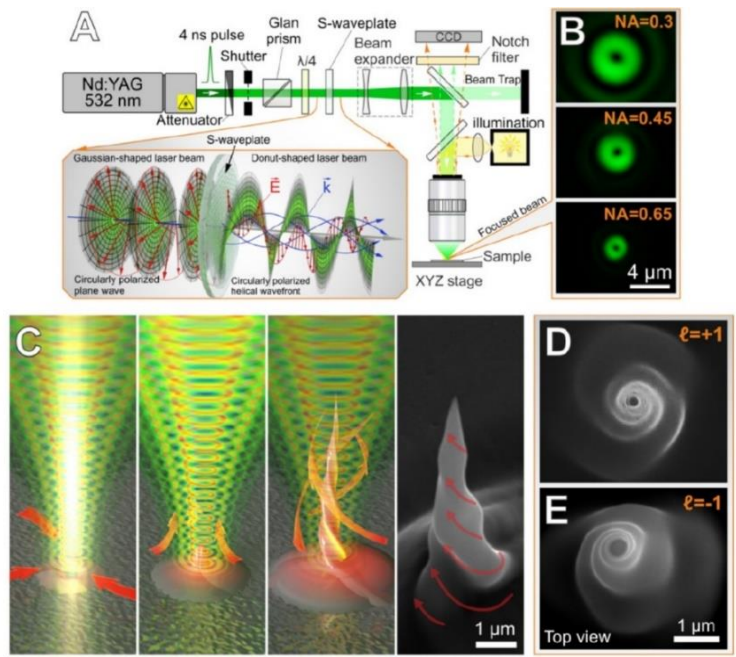

**Figure 10.** a) A scheme of the experimental setup for nanostructuring with the nanosecond vortex beams. b) Vortex-beam intensity distributions measured in the focal plane of the microscope objectives with different NAs. c) A sketch of vortex-beam-induced formation of twisted Ag nanojets. A SEM image of a twisted nanojet is shown on the right with the red arrows indicating the rotation direction. d) and e) Top-views of SEM images of twisted Ag nanojets produced under single-pulse ablation of 500-nm thick Ag film by ns vortex pulses with opposite signs of helicity. The figure is reprinted from [152], Copyright 2017, under Creative Commons Attribution. Retrieved from https://doi.org/10.1364/OE.25.010214.

One of the schemes to create vortex beams is shown in Figure 10a [152]. The material ablated with such a beam has been shown to form twisted nanostructures on metal surfaces with controlled chirality [153]. They are formed due to the involvement of ablated/molten material in a spiral motion of the vector vortex beam as illustrated in Figure 10c.

The spatial distribution of the laser fluence $\phi(r)$ of a DS beam along the radial coordinate $r$ is well described by the following equation [148,154,155]:

$$\phi(r) = 4E_0 \frac{r^2}{w_0^2} \exp(-\frac{2r^2}{w_0^2}) \qquad (1)$$

where $E_0$ is the pulse energy and $w_0$ is the waist of the corresponding Gaussian beam hosting the doughnut. According to Eq. (1), the beam fluence (intensity) is zero at the center ($r = 0$) and has a maximum at a distance $r_D = w_0/\sqrt{2}$ from the center.

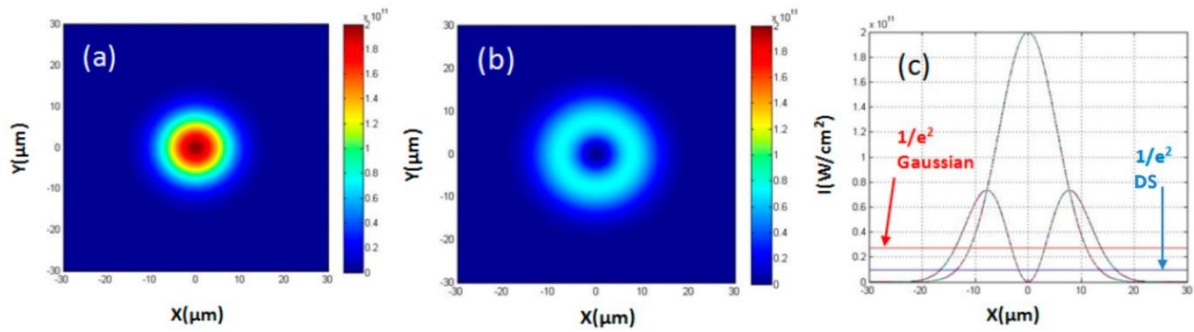

**Figure 11.** Simulated intensity distributions. a) a Gaussian beam; b) a DS beam with the same pulse energy; c) Cross-section intensity profiles of the Gaussian and DS beams. The lines show the 1/e² intensity levels for both beams. The figure is reprinted from [155] Copyright 2021, under Creative Commons Attribution. Retrieved from https://doi.org/10.3390/mi12040376.

Pulse energy of a DS beam should be ~2.7 times higher than the corresponding Gaussian beam to induce ablation if we assume that the ablation threshold is unaltered by the ring intensity distribution. Figure 11 shows simulated intensity distributions of a Gaussian and DS beam with the same pulse energy [155].

When a DS pulse is focused inside a liquid at a fluence above the liquid breakdown, a ring CB is generated, followed by a series of transient phenomena [156]. The evolution of the DS-laser-produced bubble is much more complicated than that of a typically semispherical bubble induced by a Gaussian beam. Figure 12 illustrates the early-stage dynamics of bubbles generated in water by two DS beams with 6-ns, 532-nm pulses, Figure 12a, and 150-ps, 800-nm pulses, Figure 12b. Under both irradiation conditions, the DS beam generates an annular CB and two shock waves, an outer wave moving outward and an inner wave converging toward the center. After ~50 ns, the inner wave reaches the focus, reverses direction, and triggers a secondary bubble. As this diverging wave interacts with the annular bubble, it induces a tertiary bubble cloud that vanishes within ~100 ns. Finally, the central bubble collapses after 1–2 µs, and the annular bubble fragments into cylindrical bubbles within ~100 µs. This evolution differs from Gaussian pulses in liquids when a single bubble exhibits large-amplitude oscillations and only external shock waves are emitted [157].

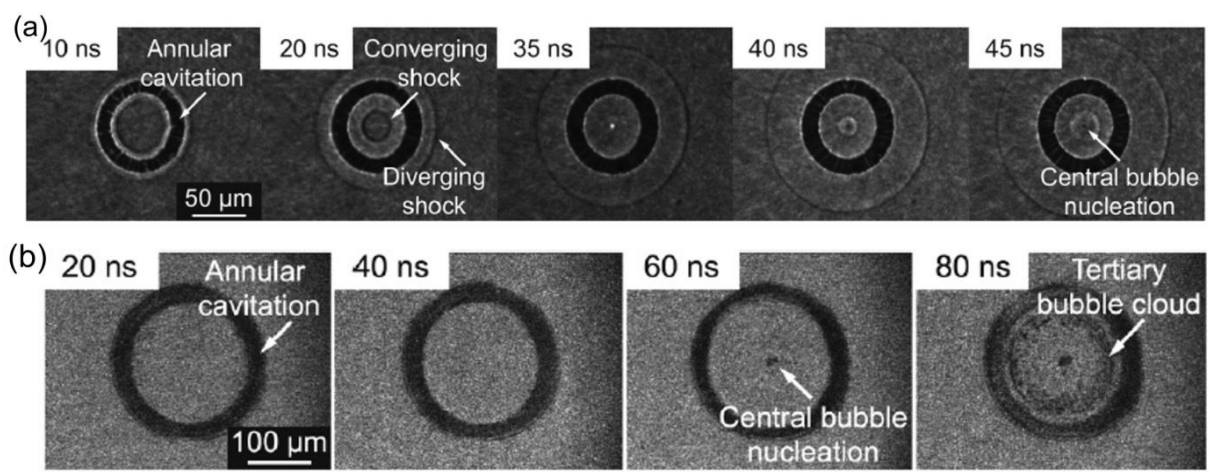

**Figure 12.** Time-resolved images of CBs in water produced by DS laser pulses at early evolution stages. a) Laser wavelength 532 nm, pulse duration 6 ns, $F_0 = 25 \, J/cm^2$, $r_D = 68 \, \mu m$. Wavelength 800 nm, pulse duration 150 ps, $E_0 = 0.5 \, mJ$, $r_D = 95 \, \mu m$. Reprinted figure with permission from[156] Copyright 2018 by the American Physical Society.

The evolution of CBs induced by DS beams in liquids can influence NP formation in PLAL by acting as nanoscale reactors. Modifying CB dynamics directly affects nucleation, offering new ways to tailor NP properties [158]. Despite the advantages of DS beams in various laser applications, their use in PLAL for NP synthesis has only recently been explored by modifying the distribution size of metal NPs [159]. Studies comparing picosecond Gaussian and DS beams on different targets (metal, oxide, alloy) showed that DS pulses significantly reduced NP size, narrowed size distribution, and improved sphericity. The NPs' size distribution shows only weak dependence on pulse energy. No significant variation was observed using 85 µJ and 217 µJ on the Au NP size distribution [159]. Instead, the observed effects are linked to the modified geometry and dynamics of the cavitation bubble under DS irradiation. The ring-shaped energy deposition alters plasma confinement and leads to cavitation bubbles with asymmetric expansion and collapse, reducing the bubble lifetime. These modified bubble dynamics influence nucleation and quenching rates, thereby limiting coalescence and favouring the formation of smaller NPs. Figure 13 compares SEM images and size distributions of Au NPs produced by PLAL in water with both laser beams. SEM analysis of Au NPs revealed that DS pulses suppressed aggregation, eliminating large particles unlike Gaussian beams, which produced size distributions with high-size tails. Additionally, DS-induced CBs had shorter lifetimes than those generated by Gaussian pulses, with the same pulse energy, possibly reducing NP aggregation. Therefore, DS laser pulses offer a promising approach for precise control over NP size and shape uniformity, although further studies are needed to elucidate the mechanisms involved.

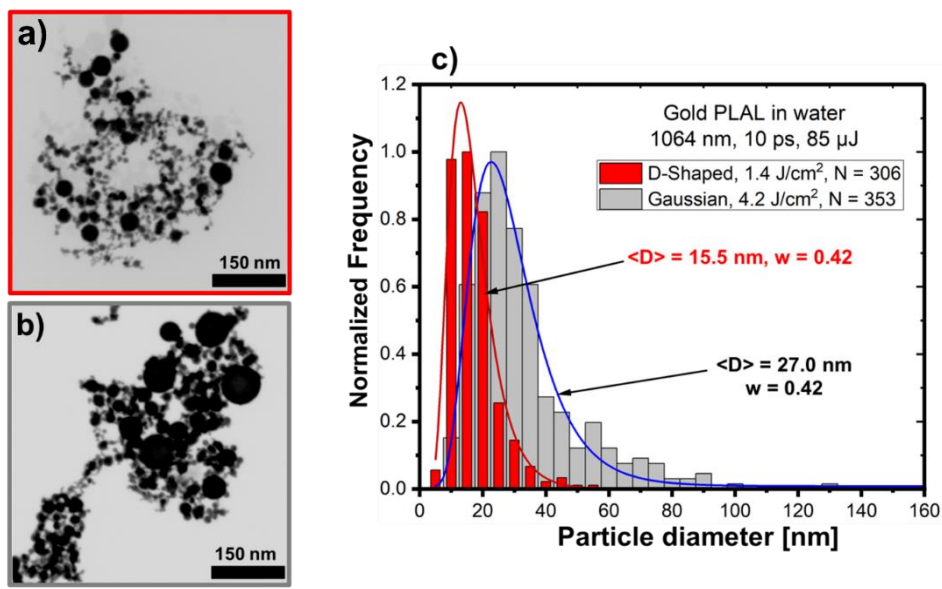

**Figure 13.** Comparison of gold NPs produced by PLAL with picosecond radially-polarized DS and linearly-polarized Gaussian laser beams at the same pulse energy of 85 µJ. a), b) SEM micrographs of NPs obtained using the DS and Gaussian beams, respectively. c) NP size distributions showing the number of NPs analyzed *N*, the median diameter *<D>*, and polydispersity index *w* values derived from the log-normal approximations (solid lines). The figure is reprinted from [159], Copyright 2025, under Creative Commons Attribution. Retrieved from https://doi.org/10.3762/bjnano.16.31.

In addition to LAL, related approaches such as laser fragmentation in liquids (LFL) can also be employed, in which preformed colloidal NPs are irradiated to induce size reduction or reshaping.

Zheng et al. showed that the use of diffractive elements/diffusers can contribute to a more efficient control of the NP size, thanks to better energy redistribution [160]. Commonly, a laser beam flat-top shape is considered to maximize laser-matter interaction. Instead, a diffused laser beam could be crucial if the process of interest has a fluence or intensity threshold. The efficiency of size-reduction of colloidal NPs was improved by a counterintuitive redistribution of laser energy, i.e., the formation of speckle patterns by incorporating a diffuser, Figure 14.

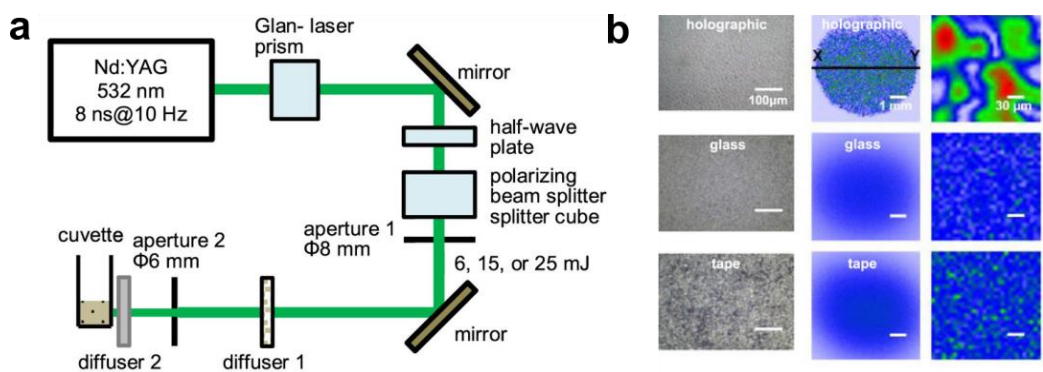

**Figure 14.** a) Experimental setup. Difusser 1: holographic. Difusser 2: glass diffuser or Scotch tape. Pulse energies of 6, 15, and 25 mJ after aperture 1 correspond to the laser fluence of 13, 34, and 56 mJ/cm² at the cuvette. b) Optical images, spatial beam profiles, and X-Y cuts of the spatial beam profiles at cuvette position through holographic and glass diffuser, and Scotch tape. Used with permission of IOP Publishing Ltd. from [160]; permission conveyed through Copyright Clearance Center, Inc.

Systematic results for Ag and Au NPs have been reported. The physical origin of the efficiency of the diffused laser beam is the redistribution of laser energy so that the local laser fluence exceeds the size-reduction threshold. Figure 15 compares the UV-Vis spectra of colloidal Ag NPs irradiated with normal and diffused beams at pulse energies of 15 and 25 mJ (fluence of $34$ and $56\ mJ/cm^2$). As seen in Figure 15a, at 15 mJ, the standard beam has minimal impact on NP size, with the surface plasmon resonance (SPR) peak at 390 nm remaining weak even after 60 minutes of irradiation. In contrast, the diffused beam at the same energy reduces NP size rapidly, and the 490 nm peak disappears within 10 minutes Figure 15b. A slight blue shift of the SPR peak between 10 and 60 minutes suggests further size reduction over time. At 25 mJ, the standard beam also induces size reduction, Figure 15c, but less effectively than the diffused beam at 15 mJ. The diffused beam at 25 mJ Figure 15d achieves even greater efficiency, further reducing NP size. Additionally, the SPR peak width narrows, indicating a more uniform size distribution. Overall, the diffused beam enhances both NP size reduction and distribution uniformity more effectively than the standard beam, showing the possibilities that spatial beam shaping can offer in NPs control.

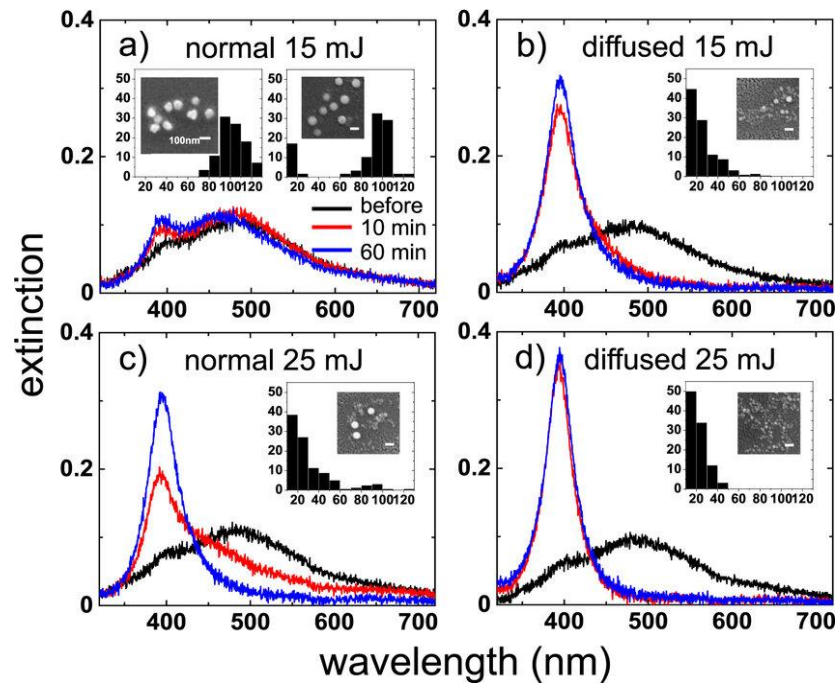

**Figure 15.** Ag NPs colloidal solutions UV-vis spectra. Without irradiation (black), after 10 min (red), and 60 min (blue) irradiation. a) Normal and b) diffused beams at 15 mJ, same for c) and d) at 25 mJ, respectively. Insets show the particle size distributions representing the diameter and SEM images. The left one in a) being those before laser irradiation while all others after 10 minutes of irradiation. Used with permission of IOP Publishing Ltd. from [160]; permission conveyed through Copyright Clearance Center, Inc.

**Bessel beams**

- Extended focal depth, stable alignment, reproducible ablation.
- Narrow NP size distributions.
- Small effective ablation area, lower productivity compared to Gaussian beams.

**Doughnut-shaped beams**

- Zero central intensity avoids overheating.
- Modified cavitation bubble dynamics reduce aggregation.
- Yield smaller, more uniform NPs (~30 nm, suppressed large-size tails).
- Require higher pulse energies (~2.7 times the Gaussian threshold).
- More complex experimental implementation.

**Diffused beams**

- Mostly employed for Laser Fragmentation in Liquids (LFL)
- Redistributes energy above fragmentation threshold.
- Enhances NP size reduction and uniformity (faster blue-shift in UV-vis spectra, narrowed SPR).
- Less control over local fluence distribution.

Spatial beam shaping directly modulates irradiance and fluence, thereby controlling bubble dynamics and NP formation. For instance, defocusing experiments demonstrate colloid concentrations up to 260 times higher when the lens is moved slightly beyond the focal plane, evidencing that maximum productivity does not always coincide with the smallest spot [59]. Reports of NP size either decreasing or increasing with defocusing can be reconciled: without power compensation, fluence drops and smaller NPs (tens of nm) with narrower distributions are obtained; with compensation, a larger ablation area produces higher concentrations and larger mean sizes [144]. More advanced shaping, such as Bessel beams, narrows Ag NP distributions (~10–20 nm) due to extended focal depth and reproducibility [143], while doughnut-shaped beams suppress aggregation of Au NPs, yielding diameters of ~30 nm versus broad, high-tail distributions under Gaussian irradiation [159]. Cylindrical focusing further enables anisotropic structures such as Ag nanoribbons up to 0.6–1 µm in length [141]. Altogether, these results highlight that spatial beam shaping offers powerful but highly parameter-dependent control: the same strategy can either fragment colloids or broaden distributions depending on local fluence, pulse duration, and liquid conditions.

# 4. Temporal Pulse Shaping in PLAL

Temporal pulse shaping is well established in laser material processing to create and modify new materials and enhance micro- and nano-machining processes [5,28].

However, PLAL innovations have been mostly focused on the evaluation of the pulse duration and repetition rate effect on the process. The range of available lasers from continuous wave (CW) to ultrashort pulses offers a broad range of parameters. CW laser irradiation generally decreases NPs production yield compared to pulsed sources for continuous large-scale production. When a continuous laser is employed, the technique is known as continuous wave laser ablation in liquid (CLAL). Since CLAL deposits energy continuously, it can lead to the suppression of the cavitation bubble and monodisperse NP size distributions[161,162]. However, the constant emission of CW lasers heats the target, inducing boiling of the surrounding liquid. The boiling liquid and the generated bubbles scatters the incident beam, which makes CLAL unfeasible for continuous or large-scale nanoparticle production[5]. Nanosecond pulses significantly reduce the thermal interaction with the liquid, but their pulse duration is longer than the typical material electron-phonon relaxation time, of the order of 0.1 - 10 ps for metals, making the thermal ablation processes dominant [163]. In the picosecond range, the pulse duration is short enough to reduce excessive heat transfer into the surrounding material but is still long enough to allow localized heating, combining photomechanical and thermal effects in the ablation mechanism. While, in femtosecond pulses, due to their ultrashort nature, the energy is deposited instantaneously into the material. This enables non-thermal ablation, where ionization and photomechanical effects dominate the material's response [164,165].

## 4.1. The role of temporal beam shaping in PLAL

In this section, we will address the impact of pulse duration on the mechanisms underlying NP synthesis within liquid media. We will consider the different phenomena from pulse emission from the laser source until NP fabrication. First, we will start with the laser's interaction with the liquid medium, alongside the nonlinear effects arising therein. Second, we will explore the energy and heat transfer to the ablation target. Finally, we will explore different ablation mechanisms relative to each distinct temporal pulse regime.

The laser pulses before reaching the liquid propagate in air. If no extreme focusing is carried out, the propagation in air will not be affected by non-linear effects, which require a high intensity, $> TW/cm^2$. The air-liquid interface represents the first difference between PLAL and material laser processing in air, the change in refractive index shifts the laser focus and introduces spherical aberration [166]. This leads to weaker focusing with higher spot sizes which reduces the fluence at the focal plane [167]. Fluence at the liquid surface can be controlled by varying the distance between the lens and the ablation system, varying the focal length, numerical aperture or processing out of focus. When the fluence on the liquid surface surpasses a specific threshold that depends on the liquid, 0.33 J/cm² for water, vaporization occurs, leading to vapor generation [59]. The distance between the lens and the liquid surface, d, to induce vaporization can be correlated with the vaporization threshold fluence of the material $\phi_{th,vap}$:

$$d = h - \omega \sqrt{\frac{\phi_0}{\phi_{th}} - 1} \qquad (2)$$

where $\phi_0$ is the focal fluence of the incident laser beam, h is the liquid layer over the target and $\omega$ is the size of the focused beam.

Intensity also plays a crucial role, defining the emergence of nonlinear interactions. The shorter the temporal duration of a pulse, the stronger the influence of nonlinear effects due to the higher intensities reached. As discussed above, non-linear effects in liquid media depend on a threshold. In this section, the same optical system and target positioning is assumed, focusing on the pulse temporal features. Consequently, the size of the processing spot is the same in every situation considered. Therefore, the dependence of the intensity is the same as the peak power, $P_{peak}(W) = \frac{E_p(J)}{\tau_{pulse}(s)}$, as it is determined by the duration of the pulse. Different temporal regimes favour the appearance of different non-linear effects. For shorter pulses, it the easier to reach higher $P_{peak}$ [5]. In femtosecond lasers, non-linear optical absorption is a prominent factor due to high $P_{peak}$ [168], giving raise to self-focusing, filamentation, and optical breakdown [169–171]. The non-linear effects reduce the energy reaching the target and affect the beam spatial and temporal profile. Picosecond lasers provide a balance between non-linear optical absorption and thermal diffusion effects [172]. Self-focusing is less intense than with femtosecond pulses, achieving effective focus without significant filamentation when operating at subcritical $P_{peak}$ [173]. Nanosecond lasers, with longer pulse duration, have lower $P_{peak}$ and thus exhibit weaker non-linear effects [174]. The longer interaction time shifts the process toward thermal effects, with substantial plasma formation occurring as the material absorbs more heat over time. Plasma screening effects become important, shielding laser energy from reaching the target. These plasma effects are less pronounced with shorter pulses, such as picosecond or femtosecond lasers given the picosecond temporal scale of plasma emergence and development, allowing higher NP yields. In summary, while femtosecond lasers are highly precise, for PLAL the strong non-linear effects limit the productivity. Nanosecond lasers exhibit thermal processes that can reduce efficiency. Picosecond lasers represent an intermediate regime where a balance between non-linear optical absorption and thermal diffusion is achieved. This balance minimizes adverse effects such as self-focusing and plasma screening, resulting in a higher NP production efficiency compared to the other pulse regimes. These general considerations strongly depend on the exact pulse duration and the power, intensity and fluence regimes employed in the experiments, being possible to find non-linear effects for high power nanosecond lasers and thermal effects for long femtosecond pulses with high power.

Each laser pulse duration offers a distinct interaction with the ablation target that differs significantly in terms of heat transfer and ablation mechanisms concerning their efficiency and accuracy in NP synthesis in liquids [175]. Pulse regime variations generate unique thermal dynamics affecting material interactions, energy absorption, and potential applications [1,176].

Femtosecond lasers facilitate swift heating that exceeds the pace of the electron-lattice coupling process. The energy deposition rate is sufficiently rapid to confine heat transfer to the target local irradiated area, thereby restricting thermal conduction to the surrounding materials [176,177]. For laser pulses in the femtosecond range, two ablation mechanisms can be differentiated. Irradiation close to the ablation threshold characterizes the "gentle" ablation phase [178]. The ablation rate is low and determined by the optical penetration depth [179]. The rapid ionization of the material leads to the ejection of charged particles due to electrostatic repulsion, named Coulomb explosion. This process occurs within a few hundred femtoseconds, indicating a swift breakdown of material under the laser influence. Besides, at lower fluences, photomechanical spallation occurs, which involves moderate temperature and pressure, removing material with minimal vaporization [180]. Surface ripples appear and the machining area remains relatively smooth. At higher fluences, in the "strong" ablation regime, phase explosion becomes significant, for Au $\phi_{th}^s = 0.86\ J/cm^2$ [181], resulting in a rougher surface and an increased ablation rate [180].

Therefore, the ablated volume $\Delta V$ as a function of fluence is related to the optical penetration depth $\delta$ and the electron-thermal penetration depth $l$ due to electron heat conduction. Considering that the Gaussian beam fluence distribution follows $\phi = \phi_0 \exp(-r^2/\omega^2)$, with $\omega$ being the radius beam and $\phi_0 = E_0/\pi\omega^2$ the peak fluence:

$$\Delta V = \frac{E_0}{2\phi} \ln\left(\frac{\phi}{\phi_{th,\delta}}\right) \left[ A\delta \ln\left(\frac{\phi}{\phi_{th,\delta}}\right) + Bl \ln\left(\frac{\phi}{\phi_{th,l}}\right) \right] \qquad (3)$$

for $\phi_{th}$ the threshold fluence, $E_0$ the absorbed energy and with $\phi_{th,\delta}$ and $\phi_{th,l}$ the threshold fluences regarding the two different penetration depths [20]. The constants A and B in Equation (3) are introduced from the experimental data.

In comparison, the picosecond pulses duration is in the range of the electron-phonon coupling time of common materials, 1-5 ps, resulting in a local heating process [182]. This energy transfer creates a state of nonthermal equilibrium, where the electron temperature rises quickly before the lattice (atomic structure) can respond, leading to localized heating. The energy is delivered quickly enough to limit significant thermal effects, yet there is still thermal diffusion [177]. For pulses longer than 1 ps, the ablation regime determined by the optical penetration depth is not observed [179]. In the nanosecond regime, the energy is deposited over an extended period, and heat has time to spread through the material. Prolonged energy deposition enables heat conduction, causing a significant increase in temperature over a wider area of the material, leading to gradual heating, melting, and vaporization [183]. Since the pulse duration exceeds the thermalization time of most metals [163], the nanosecond regime typically results in thermal effects that impact the irradiated and the surrounding material [175]. It also involves photothermal effects that result in vaporization, as well as photoionization, which facilitates plasma generation, ejecting material from the target surface [184]. At high laser fluences, the phase explosion mechanism can occur. The comparison of the different pulse durations, typical parameters, and the associated PLAL mechanisms are summarized in Table 1.

**Table 1:** Comparison of temporal pulsed laser ranges in PLAL. Non-linear effects thresholds correspond to measurements and theoretical calculations performed for the laser interaction with water.

| | Pulse Duration | Pulse Energy | Irradiance/Intensity | Laser-target interaction | Ablation Mechanism | Non-lineal effects threshold |
|---|---|---|---|---|---|---|
| **Nanosecond** | $10^{-9}$ s | 10–500 mJ | 30GW/cm$^2$-1.5TW/cm$^2$ | Heat conduction increase temperature | Heating Melting Evaporation | **Self-focusing** 0.01TW/cm$^2$ |
| **Picosecond** | $10^{-12}$ s | 10µJ-10mJ | 40GW/cm$^2$-40TW/cm$^2$ | Localized heating thermal diffusion | Phase explosion | **Filamentation** 1TW/cm$^2$ |
| **Femtosecond** | $10^{-15}$ s | 1µJ-10mJ | 0.4TW/cm$^2$-4PW/cm$^2$ | Quickly electron-lattice coupling | Multiphoton ionization/ Phase explosion | **Optical breakdown** 10TW/cm$^2$ |

To obtain the values displayed in Table 1, the same optical system is assumed consisting of a processing area of 100 µm beam diameter at the focal plane. The pulse durations considered for calculation have been chosen as the most common for each range, 4ns, 3ps and 30 fs.

The role of the laser repetition rate in NP synthesis represents a key area of research within PLAL. The interaction between a laser and a solid target immersed in a liquid generates CBs during the plasma cooling [18]. Understanding the dynamics of CB formation and its interaction with subsequent laser pulses is essential for optimizing NP yield. Otherwise, the productivity for long-term PLAL experiments will be drastically reduced [185,186]. Therefore, increasing productivity requires higher repetition rates, to spatially bypass the bubble [53]. This strategy involves matching the scanning velocity to the bubble size and the laser frequency according to the relationship $v = r \times f$, where $v$ is the scanning velocity, $r$ is the bubble radius and $f$ is the repetition rate. Advanced laser systems, operating at MHz frequencies, coupled with high-speed scanners, can reach the highest NP yields [129]. Indeed, the current maximum PLAL productivity was achieved by a fast polygon scanner (500 m/s) to spatially bypass the CB, achieving an ablation rate of 8.3 g/h for platinum [54]. Repetition rate significantly affects energy accumulation and heat management in the target material [178,187], directly impacting the NP size, shape, and productivity. Studies using femtosecond lasers have highlighted that increasing the repetition rate tends to reduce the mean NP diameter. When the repetition rate is increased pulse overlap increases proportionally, given that the scanning speed is kept constant. For increasing pulse overlap the freshly generated NPs further absorb part of the incoming laser energy from the subsequent pulses. This secondary irradiation can modify their crystallinity and typically results in smaller particle sizes due to laser fragmentation. The phenomenon is continuous and gradual, as the repetition rate increases, the cumulative overlap between successive pulses becomes more pronounced, leading to a progressive reduction in the NPs size [188]. Upon irradiation with a picosecond laser, the CB in water typically exhibits a lifetime of 100 µs and a diameter of 100 µm. [53,189]. For nanosecond pulses, it has been reported that the lifetime of CBs can extend to approximately 250 µs, with a maximal radius reaching 1.4 mm. [22]. Nonetheless, the CB features are associated with the structural and morphological characteristics of the irradiated sample [190], the fluence applied, and the solvent [10,22,130].

### 4.2. Temporal pulse shaping innovations in PLAL

The relevance of pulse shaping approaches in PLAL can be first described by the time resolved pump-probe measurements that allow to evaluate the early dynamics of PLAL with picosecond and even femtosecond resolution. Those initial features are key to understand the influence of the pulse duration on PLAL and how the interpulse delay in a double pulse configuration influences the NP formation mechanisms.

### 4.2.1. Pump-Probe Microscopy in PLAL

Ultra-fast Pump-Probe Microscopy (PPM) is a key technique for the understanding of processes occurring during PLAL with very high temporal resolution reaching the femtosecond scale. Particularly, for providing extensive temporal information about the early moments involving the spallation layer and the initial stages of NP formation. PPM has been employed to monitor NP generation and microparticle fragmentation in liquid with a temporal resolution of 500 fs and a spatial resolution of 600 nm [191]. The whole ablation dynamics of gold irradiated with a peak fluence of $\phi_0 = 1.5\phi_{th}$ in air and water was recorded for delay times $\Delta t$ ranging from ps to µs [192,193]. The $\phi_{th}$ denotes the single-pulse ablation threshold fluence [18]. The results of the transient reflectivity changes $\Delta R/R_0$, showed that the ablation dynamics can be divided into seven temporal domains, Figure 16a. PLAL dynamics differ significantly compared to laser ablation in air, following the labels included in Figure 16b and starting from domain ②. The water layer promotes high-density plasma generation by an interplay of thermionic emission from the hot Au surface and optical breakdown near it. Following plasma dilution in domain ③, the target ablation in regime ④ experiences spallation or phase explosion, depending on the incident peak fluence. In domain ⑤, it could be shown that NPs already form several 100 ps after pump-pulse impact. After NPs are generated, a shock wave is emitted, and a CB is consequently formed at $\Delta t$ = 1 ns. Therefore, the shock wave and CB dynamics govern the remaining temporal range in domain ⑥, leading to NPs and microbubbles dispersion in water within domain ⑦.

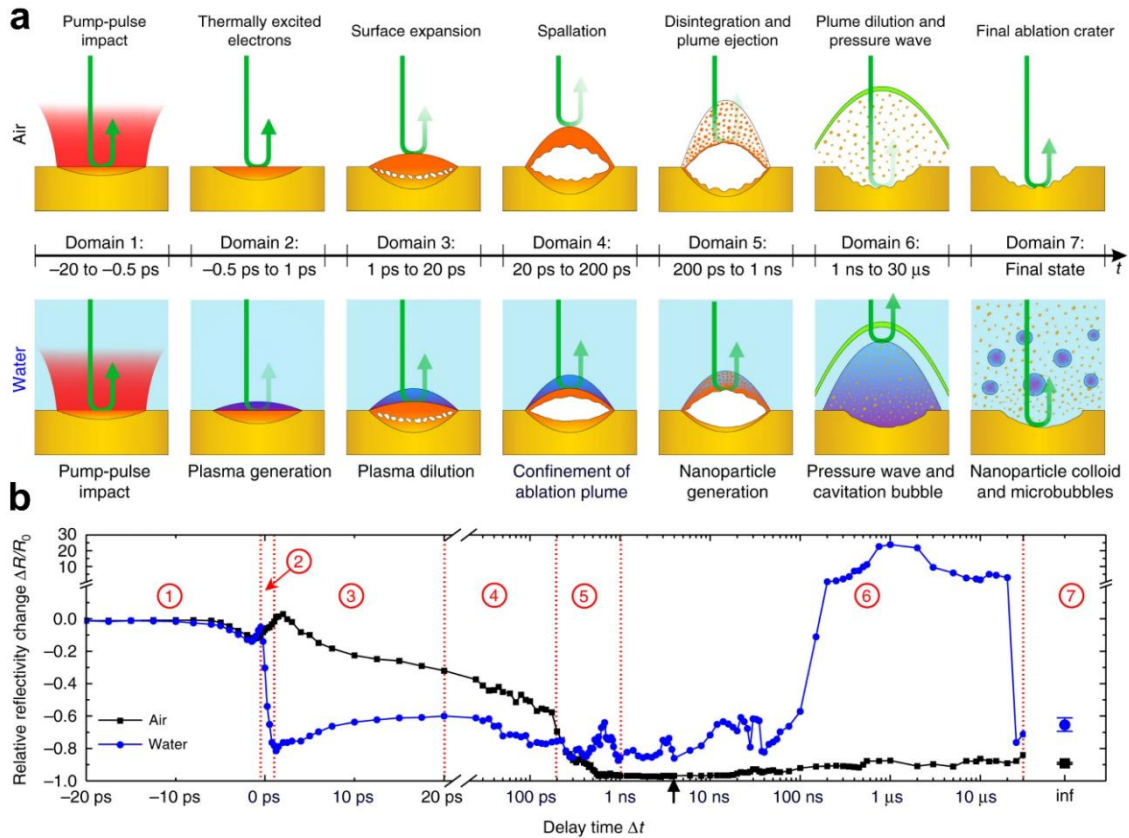

**Figure 16.** Comparison of pump-prove laser ablation experiments of gold immersed in air and water irradiated with an incident peak fluence of $Φ_0 = 1.5·Φ_{thr}$. a) Schematic depiction of the ablation dynamics occurring within the seven temporal domains. b) Transient relative reflectivity change (ΔR/R0) for gold in air and water, with seven temporal domains indicated by red encircled numbers. The figure is reprinted from [18], Copyright 2022, under Creative Commons Attribution. Retrieved from https://doi.org/10.1038/s41377-022-00751-6.

PLAL dynamics span 9 orders of magnitude in time, ranging from plasma generation on a ps timescale to CB collapse on a μs timescale. NP generation occurs on sub-ns timescales, as predicted by computational studies [194,195]. Furthermore, the onset of the CB could be identified to occur approximately 1 ns after the pulse impact [9,26].

Furthermore, in the femtosecond range, the ablation process of iron in air and water was investigated by PPM for fluences of $0.5\,J/cm^2$ and $2\,J/cm^2$ [196], as depicted in

Figure 17a. As noted above, measurements of the relative change in reflectivity suggest that the surrounding liquid has a significant impact on the ablation process, Figure 17b. In the heating phase, 10 ps after laser impact, there are no significant changes in reflectivity. Once this period has elapsed, the reflectivity does not decrease due to the scattering and absorption of the confinement of the hot target material by the liquid. In contrast to ablation in air, reflectivity increases due to the hot dense metal layer at the plume-liquid interface, as shown in Figure 17c.

The ultrafast temporal resolution provided by PPM enables a detailed, real-time understanding of the dynamics of plasma generation, NP formation, and shockwave propagation. PPM has also been successfully applied to probe NP generation by microparticle fragmentation in liquid [191]. PPM experiments also shed light on the influence of different confining liquids on the ablation process [196]. These studies enhance the control and optimization of PLAL by controlling the ablation mechanism and providing direct information of the most suitable laser parameters, including fluence, pulse duration and inter-pulse distance. Moreover, these results highlight the potential for temporal pulse shaping to influence and control the outcomes of PLAL.

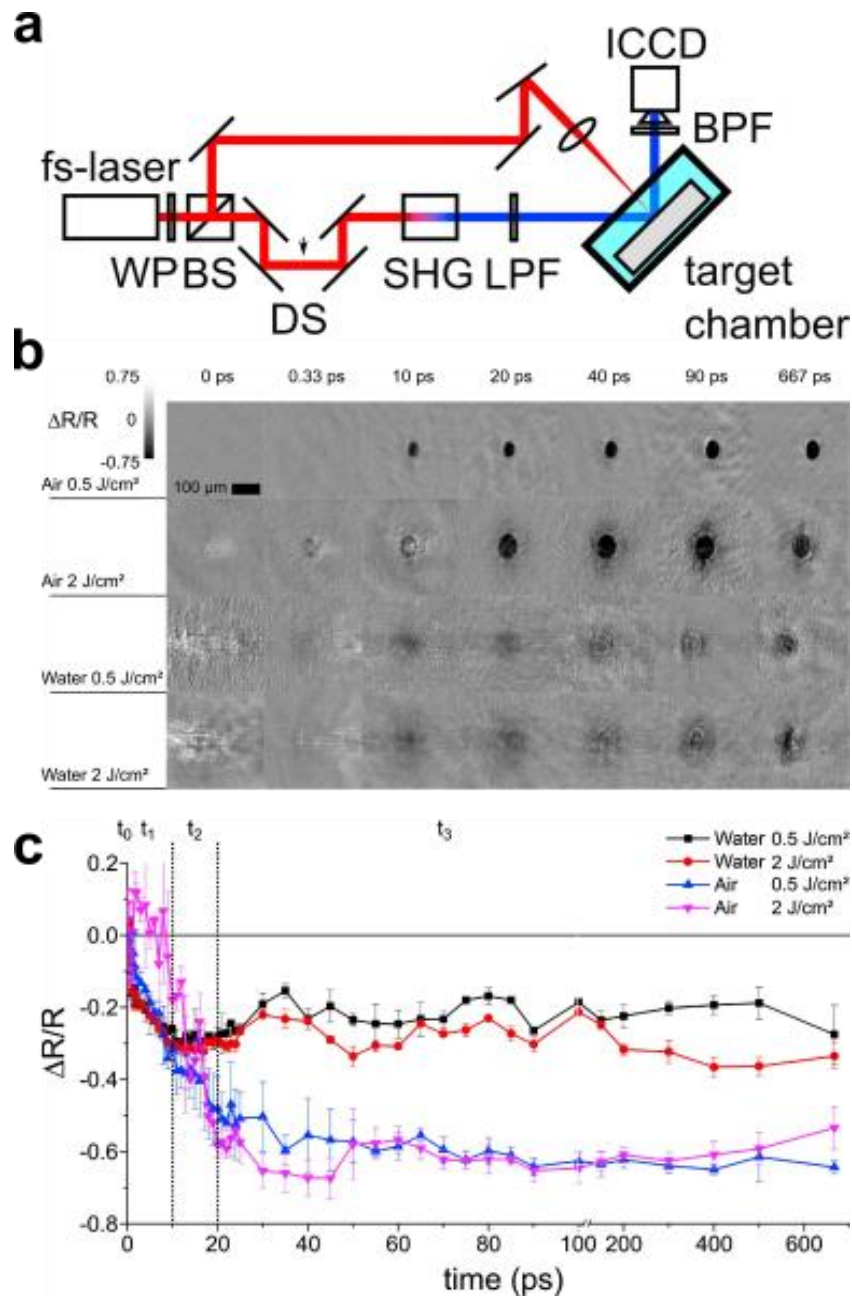

**Figure 17.** a) Schematic depiction of the Pump-Prove experimental setup. b) Microscopy images of the time-resolved ablation process in air and water for $0.5\,J/cm^2$ and $2\,J/cm^2$ fluences. c) Relative reflectivity changes mean-value for ablation in air and in water. Reprinted from [196] Copyright 2018, with permission from Elsevier.

A direct comparison between ablation in air and liquids highlights clear differences in thresholds, yields, and mechanisms. For gold, representative ablation thresholds under ultrashort irradiation are typically in the range of $\sim 1-2\,J\cdot cm^{-2}$ incident fluence in air (e.g., 1.3 $J \cdot cm^{-2}$ at 124 fs, 800 nm) [197]. For picosecond pulses, Spellauge et al. reported ~1.4 $J \cdot cm^{-2}$ in air and ~2.1 $J \cdot cm^{-2}$ in water, though the corresponding absorbed thresholds are comparable (~0.04 and ~0.06 $J \cdot cm^{-2}$) [18]. In the nanosecond regime, thresholds in water are often higher than in air by ~1.2 − 1.6 factor [198]. Regarding NPs yield, Dittrich et al. reported ~5 $\mu g \cdot (W \cdot s)^{-1}$ for 3 picosecond laser ablation of gold in water, lower than in air ~40 $\mu g \cdot (W \cdot s)^{-1}$ [137]. Streubel et al. achieved up to ~4 $g \cdot h^{-1}$ and 8.3 $g \cdot h^{-1}$ production Au and Pt NPs in water using high-power picosecond pulses with fast scanning [53,54]; simpler setups typically reach tens of $mg \cdot h^{-1}$. It illustrates the strong influence of the laser and liquid parameters on NP yield. Mechanistically, ablation in air is governed by plasma expansion and shockwave emission, while in liquids the formation, oscillation, and collapse of cavitation bubbles influences NP nucleation and growth. Absolute threshold and yield values strongly depend on multiple parameters, including pulse duration, laser wavelength and power, liquid and layer, surface preparation of the target, and measurement methodology. Therefore, reported numbers may vary significantly across studies and should be interpreted as representative ranges rather than definitive constants.

### 4.2.2. Double-pulse PLAL

The control of the inter-pulse delay represents an approach to tune the dynamics of PLAL by modifying with a second pulse the ablation mechanisms ongoing after the first pulse. The PPM measurements provide further insights into the effect of the second pulse depending on the delay [18]. Indeed, different temporal delay regimes have been studied using double-pulse configurations in PLAL, shedding light on the interplay between pulse delay and NP formation and showing modifications of the productivity and NP size distribution [199–204].

Double-pulse laser ablation can be divided into two distinct regimes. In the first regime, the interpulse delay is approximately below 2 ns. The second pulse arrives before the CB is fully formed, and can interact with the spallation layer, the plasma or the early CB, influencing the NP formation process [18,174]. Double pulse PLAL Au experiments in the sub-nanosecond regime (300 ps to 1200 ps) have been performed with a 1064 nm, 10 ps, 100 kHz, 10 W laser, demonstrating that the NP size distribution is modified [205]. A bimodality reduction of up to (9 ± 1) wt% of the large NP fraction was achieved with an interpulse delay of 600 ps, as shown in Figure 18a. Productivity measurements as a function of the pulse delay confirm a reduction at 600 ps. Both findings indicate that the second pulse reaching the target at that pulse delay is likely to interact with the emerging spallation layer generated by the first pulse.

In the second regime, the interpulse delay exceeds 2 ns, resulting in the irradiation of the CB, hence influencing the CB and NP coalescence and growth [205]. Double-pulse PLAL was used for the laser synthesis of Si NPs in ethanol, indicating an enhancement of NP productivity with better control over the NP size distribution [204]. For longer delays, it has been proved experimentally that it is possible to reduce the NP size distribution by controlling the temporal separation of the double pulse in the µs scale [206]. The second pulse interacts with the first expanding and collapsing CB, with a time scale of 0-200 µs [206], as shown in Figure 18b, reducing drastically the NP diameter. However, simulations developed for the temporal evolution of material ejected in PLAL, Figure 18c and 18d, indicate that material ejection can be more effectively influenced in a time scale of nanoseconds (first regimen), i.e. at the time scale before the CB is formed [194].

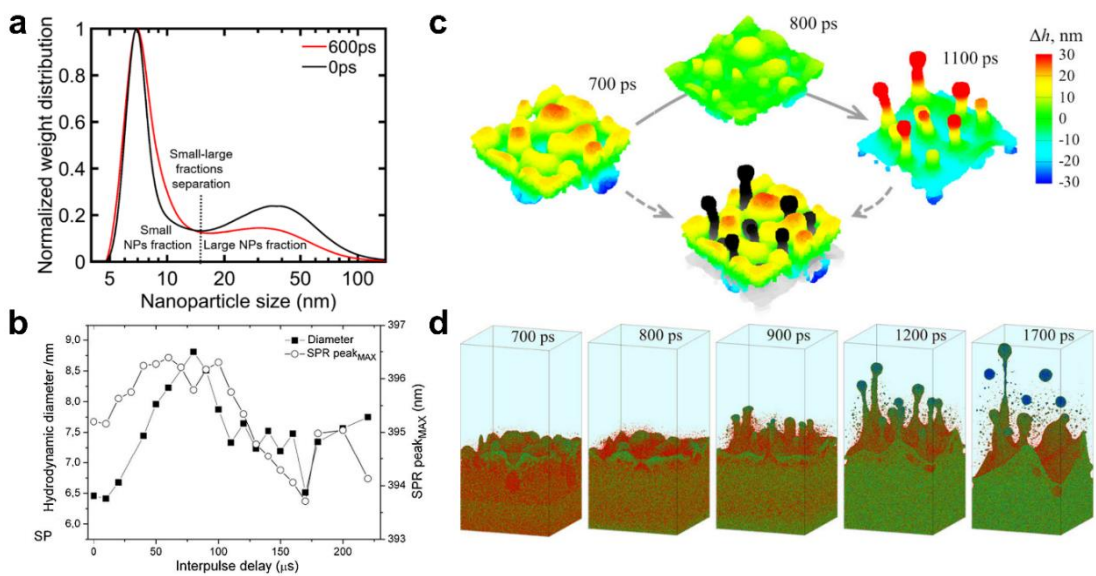

**Figure 18.** a) Normalized NP weight distribution as a function of the particle diameter for double-pulse delay times of 0 ps (black line) and 600 ps (red line). The vertical red dotted line at 16 nm shows the selected value that separates the small NP fraction from the large NP fraction. Used with permission of IOP Publishing Ltd. from [205]; permission conveyed through Copyright Clearance Center, Inc. b) Hydrodynamic diameter of Ag NPs prepared by double-pulse PLAL as a function of interpulse delay in the µs-regime. Used with permission of Royal Society of Chemistry from [206]; permission conveyed through Copyright Clearance Center, Inc. c) Surface reconstruction in the initial ns time scale after single pulse irradiation of a silver metal target for the hot metal layer−water interface generated. d) Simulated evolution of the hot metal layer−water interface predicted a bulk Ag target with a subsurface void irradiated in water. The figures c) and d) are reprinted from [194], Copyright 2017, under Creative Commons Attribution. Retrieved from https://doi.org/10.1021/acs.jpcc.7b02301.

The physical origin of the modality change lies in the evolving state of the target–liquid interface after the first pulse. For interpulse delays below ~2 ns, the surface remains in a highly non-equilibrium condition: a molten layer persists, dense plasma clouds are present near the target, and cavitation bubble growth is still incipient. Under these conditions, it is hypothesized that for a specific inter-pulse delay that it is material dependent, 600 ps for Au, the second pulse couples into the spallation layer, interacting with it and reducing the formation of larger NPs [205]. In contrast, for delays exceeding ~2 ns, the target surface has already cooled and partially re-solidified, and the dominant structure is the CB produced by the first pulse. The second pulse therefore interacts primarily with this bubble, changing its expansion and collapse dynamics, which govern NP aggregation and growth. Consequently, shorter delays mainly influence direct target modification, the spallation layer, and the initial material ejection dynamics. While for longer delays, the second pulse interacts with the CB, influencing NP nucleation and formation. This mechanistic view explains the experimentally observed transition between the two regimes.

In addition to being able to temporarily adjust the interpulse delay, double pulse PLAL experiments allow to adjust the lateral separation of synchronous pulses. When the separation between simultaneously irradiated spots is considerably larger than the maximum CB size, the ablation events can be considered independent, and the sizes of the produced NPs are identical to those obtained with a single beam [55,56]. However, if the spot separation and bubble size are comparable, the neighboring bubbles start to interact resulting in a modification of the bubbles dynamics. This technique, called synchronous-double-pulse PLAL, has been used to understand bubble interaction effects on the NP size [207]. The dynamics of closely positioned bubbles were investigated for the case of dual-beam PLAL of gold and yttrium aluminium garnet (YAG) in water [208]. Figure 19a shows selected shadowgraph images of the interacting bubbles during their evolution at different inter-pulse distances $\Delta x$. Synchronised bubble pairs are generated with the same pulse energy ($E_p$). At a small spot separation, 2 times the maximum bubble height ($\approx 2H_{max}$), an internal wall between two adjacent bubbles is formed followed by bubble merging with the formation of a larger combined bubble. The bubble interaction dynamics essentially affect the NP formation process, significantly increasing NP hydrodynamic diameters by a factor of up to 3.7, as shown in Figure 19b. In contrast, bubbles separated at large distances of $4H_{max}$, result in NP size distributions like those produced by single pulse experiments (0 separation between synchronous beams). The merging and collapse of these bubbles lead to larger NP sizes, indicating that bubble dynamics influence the particle size distribution, Figure 19c-e. Therefore, the variation of the inter-pulse lateral distance in synchronous-double-pulse PLAL allows to control the NP size distribution. This technique represents an alternative approach to tune the size of colloidal NPs without affecting their purity [208].

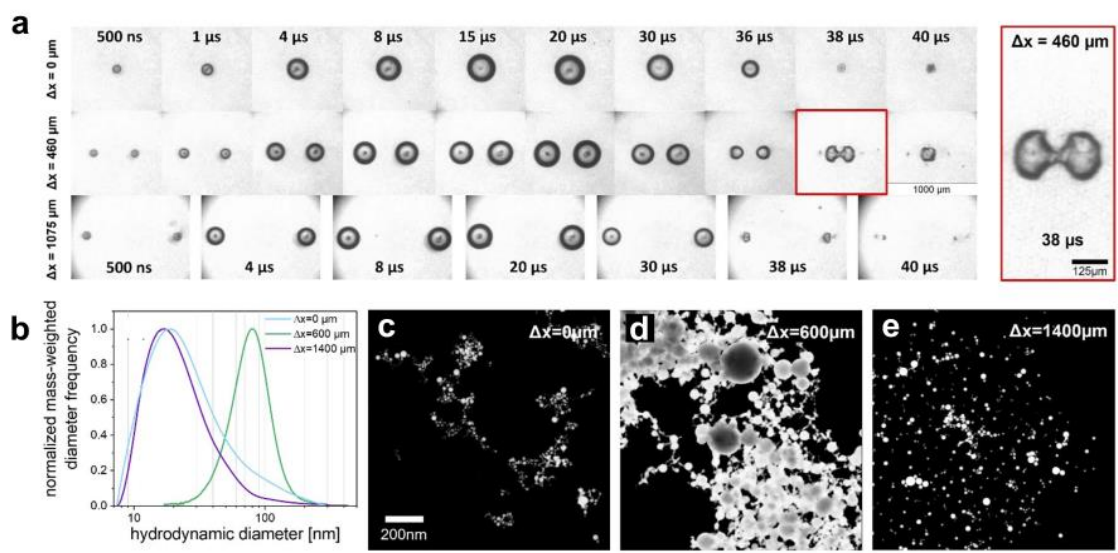

**Figure 19.** a) Shadowgraph images of CBs produced by ps dual-beam PLAL of YAG in water at the energy in each pulse of 470 µJ and different inter-pulse distances at $\Delta x = 0$, 460, and 1075 µm. The 1000 µm scale bar is the same for all image series. Has been highlighted the magnifications of the bubble pair images at $2H_{max}$. b) Hydrodynamic mass-weighted particle diameter distributions measured by analytical disk centrifuge for Au NPs. c) – e) HAADF-STEM images of Au NPs synthesized for Δx = 0, 600, and 1400 µm inter-pulse distances. The scale bar is the same for all images. The figure is reprinted from [207], Copyright 2023, under Creative Commons Attribution. Retrieved from https://doi.org/10.1364/PRJ.498204.

- Temporal delay between pulses can influence PLAL dynamics, improving ablation efficiency or modifying the resulting NP size distribution.

- Can significantly boost NP yield (up to a factor 2 compared to single-pulse) and improve energy utilization for the optimum short delays in the ps range.
- Double pulse delays in the ns and µs regime can influence cavitation bubble formation and dynamics, tuning the NP size reduction but decreasing the efficiency.
- Synchronous double pulse PLAL with controlled spatial separation can tune NP size distribution by adjusting the interpulse distance, influencing the cavitation bubble dynamics.
- Optimal separation of 2 times the maximum cavitation bubble radius leads to an increase of the NP hydrodynamic size by a factor of up to 3.7 for Au and YAG.
- The increased system complexity requires precise spatial and temporal control between pulses.

### 4.2.3. Pulse duration effect on PLAL ablation rate and nanoparticle yield

In the previous sections different approaches to control the NP size distribution have been described. However, in every case the pulse duration was kept constant, and the nanoparticle yield was kept constant compared to single pulse experiments or even reduced. Hence, ablation efficiency experiments are key for understanding the optimum laser pulse duration for each material. These experiments characterize the material removal and NP formation, providing insights into how pulse duration influences ablation efficiency and NP production.

Ablation efficiency exhibits a decline as the pulse duration increases beyond the picosecond range, shown in Figure 20, largely due to plasma shielding effects and enhanced heat conduction losses, which limit material removal efficiency and precision. Operating in the ps-pulse regime mitigates non-lineal effects, typically observed at high power levels reached for femtosecond pulses. When subcritical conditions are maintained, picosecond pulses enable stable energy deposition, promoting a more controlled and efficient NP synthesis process [171].

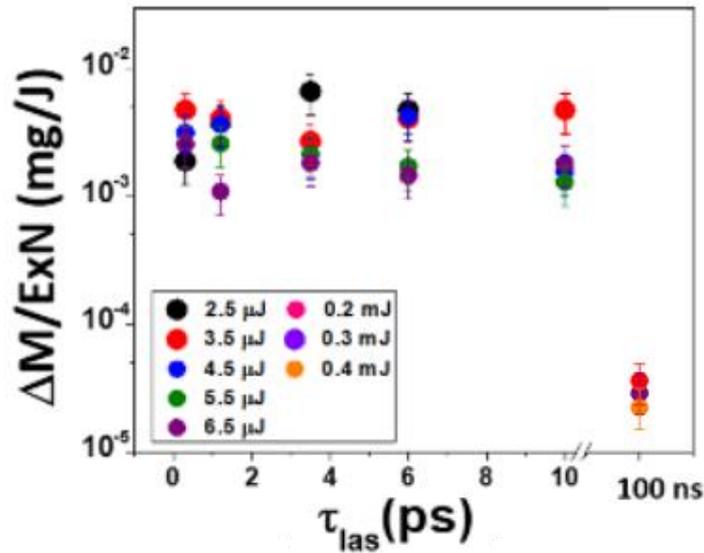

**Figure 20.** Dependence of mass loss on the pulse duration for different pulse energy and 85 nm film thickness. Used with permission of IOP Publishing Ltd. from [209]; permission conveyed through Copyright Clearance Center, Inc.

As shown in Figures 21a and 21b, the ablation in air and water results show opposite trends in terms of pulse duration. For ablation in air, the femtosecond range is optimal, due to the high peak power, increasing the production of NPs with increasing fluence. In water, the highest ablation volume and crater depth is found for picosecond pulses. The femtosecond reduced yield for PLAL is associated to the non-linear effects in the liquid and their corresponding energy losses and beam profile disturbance. Figure 21c depicts how the ablation crater in water increases considerably when using a laser in the picosecond range, 10 ps, compared to a sub-picosecond laser, 300 fs.

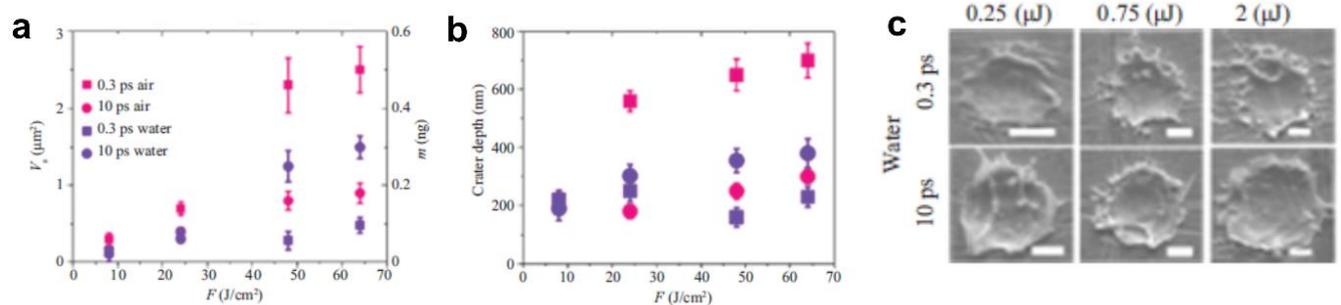

**Figure 21.** Pulse width dependence with pulse energy for a) ablated mass and b) crater depth in water. c) Crater shape after ablation. Scale bar: 1µm. The figure is reprinted from [171].

In addition, the ablation threshold and pulse duration relationship have been studied in gold films immersed in water by varying the target thickness [209]. Thickness and pulse duration influence the energy required for effective material removal. For 175 and 330 nm thicknesses, the ablation threshold decreases as pulse duration increases. Longer pulse durations, in the picosecond range (10 ps), allow for greater energy deposition depth in the electron-phonon coupling time regime. In contrast, thinner films, 85 nm, show relatively stable ablation thresholds across varying pulse durations, as shown in Figure 22a. For thin films, the ablation process is less influenced by thermal diffusion and more dependent on rapid energy transfer. The energy required for ablation remains relatively constant because heat is more effectively dissipated to the surrounding liquid compared to thicker films. In air, mass loss is more efficient for shorter pulse durations (sub-picosecond range, 300 fs) due to the reduced thermal diffusion effects [209]. Nevertheless, in water, a longer pulse duration, between 2 – 6 ps, increases crater depth up to 300 - 400 nm, as demonstrated in Figures 22b and 22c. This is attributed to the decrease in peak power, which reduces the influence of non-linear interactions with the liquid, maximizing the energy reaching the target surface. Overall, the morphology of the craters is influenced by the interplay between the pulse duration and the resulting thermal and nonlinear effects during the ablation process [169]. In the nanosecond range, crater depth and volume decrease, shifting from precision ablation to greater thermal interaction and reduced efficiency. Nanosecond pulses produce broader, less defined craters with increased convexity, exceeding the beam diameter at higher energies. In this range, plasma formation, shielding, and scattered energy reduce ablation efficiency compared to shorter pulses.

Prolonged interaction decreases particle size and uniformity control, enabling mass removal under a well-established plasma scaling relationship [171].

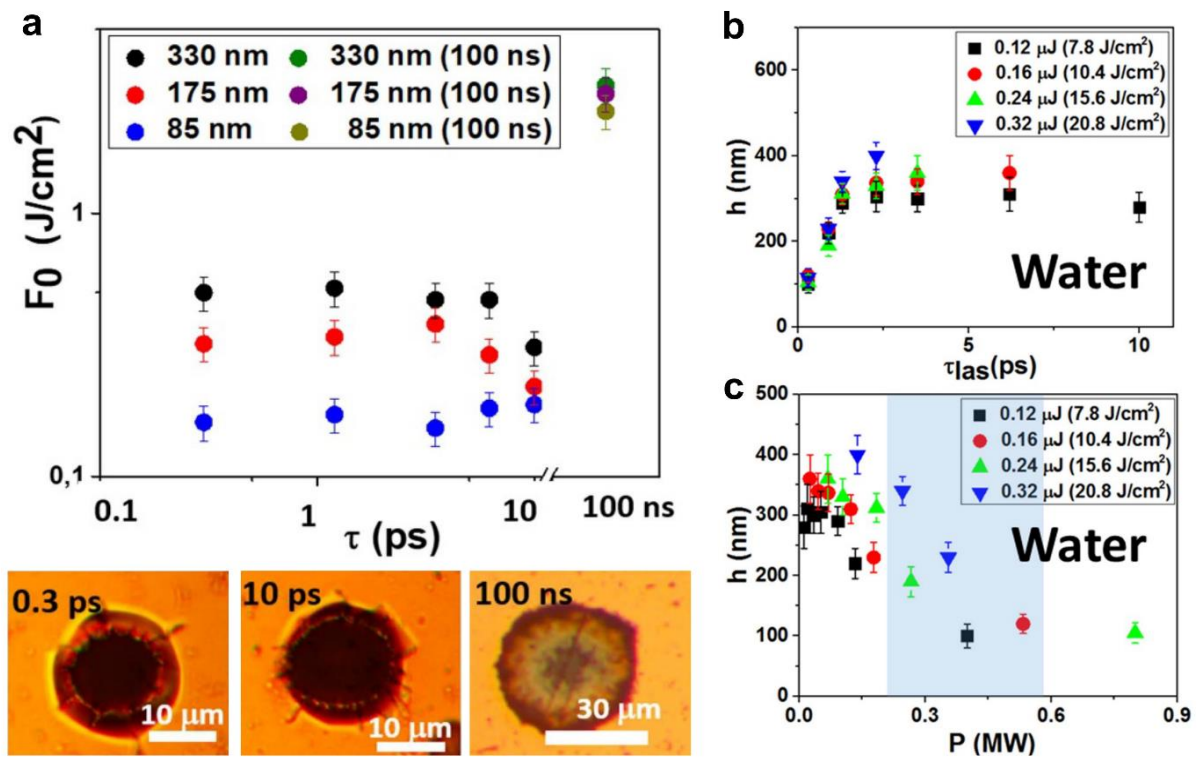

**Figure 22.** a) Threshold fluences dependence on the pulse duration; insets; optical images of single shot puls craters on 330 nm thickness fils (E = 4.5µJ for 0.3, 10 ps; 0.5mJ for 100 ns). Used with permission of IOP Publishing Ltd. from [209]; permission conveyed through Copyright Clearance Center, Inc. b) Dependences of the crater depth on different pulse durations at fixed energy values in water. c) Ablation depth versus peak power for ablation in water. Used with permission of IOP Publishing Ltd. from [169]; permission conveyed through Copyright Clearance Center, Inc.

- Fs-ps pulses lead to higher peak power, efficient ablation, reduced thermal effects, smaller and more uniform NPs.
- Fs pulses minimize the heat-affected zone.
- Fs pulses can trigger nonlinear effects like filamentation and optical breakdown that cause energy losses and lower productivity.
- Ns pulses increase thermal effects resulting in broader size distributions.
- Ps pulses are generally more productive due to higher ablation rates and lower energy losses in the liquid.

Experimental evidence shows that pulse duration critically governs ablation mechanisms, NP yield, and size control in PLAL. CW irradiation leads to inefficient high scale NP generation due to continuous heating, liquid boiling, and beam scattering. Nanosecond pulses favour thermal ablation and plasma shielding, producing wide, thermally affected craters and lower yields[209]. Picosecond pulses provide an optimal balance between nonlinear absorption and thermal diffusion, maximizing ablation volume (crater depths ≈300–400 nm) and NP yield, with record productivities of 8.3 g·h⁻¹ reported for Pt using MHz repetition rates and high-speed scanning (500 m·s⁻¹) [54]. Femtosecond pulses enable highly localized, non-thermal ablation but suffer from strong nonlinear effects (self-focusing, filamentation), reducing productivity compared to picosecond regimes [171]. Higher repetition rates (≥ 200 kHz) decrease mean NP size via secondary irradiation, while double-pulse PLAL allows additional control: delays of 600 ps reduce the large-NP fraction by ≈9 wt% [205], and µs-scale delays tune cavitation bubble dynamics, further narrowing size distributions [206].

# 5. Spatiotemporal Beam Shaping Techniques in PLAL

The techniques explored in the previous sections propose the incorporation of temporal or spatial beam shaping techniques to the standard PLAL methodology to upscale NP production or control the size distribution obtained. However, there are techniques that modify the beam both spatially and temporally to combine the benefits of both approaches for PLAL.

## 5.1. Simultaneous spatial and temporal focusing femtosecond PLAL

The difficulties related to nonlinear interactions of the femtosecond pulses with the liquid mentioned in the previous sections limit their employment in PLAL when high productivity is required. To address this limitation and reduce the nonlinear interactions, a simultaneous spatial and temporal focusing (SSTF) setup was proposed. The key idea of SSTF is to employ a diffractive grating to add a spatial chirp to the femtosecond beam, so the different spectral components are separated and only recombine at the spatial focus of the objective lens. Thus, the temporal pulse width becomes a function of the distance while propagating from the lens to the target, with the shortest pulse width recovered at the focal spot. The use of femtosecond lasers in PLAL presents a key limitation related to energy attenuation within the liquid medium [1]. The extremely high peak power of ultrashort pulses often induces nonlinear optical phenomena, such as filamentation and optical breakdown, prior to reaching the target surface, which significantly decreases ablation efficiency. The SSTF configuration effectively mitigates these drawbacks by confining the peak intensity to a narrow focal region. In this setup, the pulse duration lengthens away from the focus, suppressing nonlinear propagation effects and thereby enhancing the effective energy transfer to the target, Figure 23d.

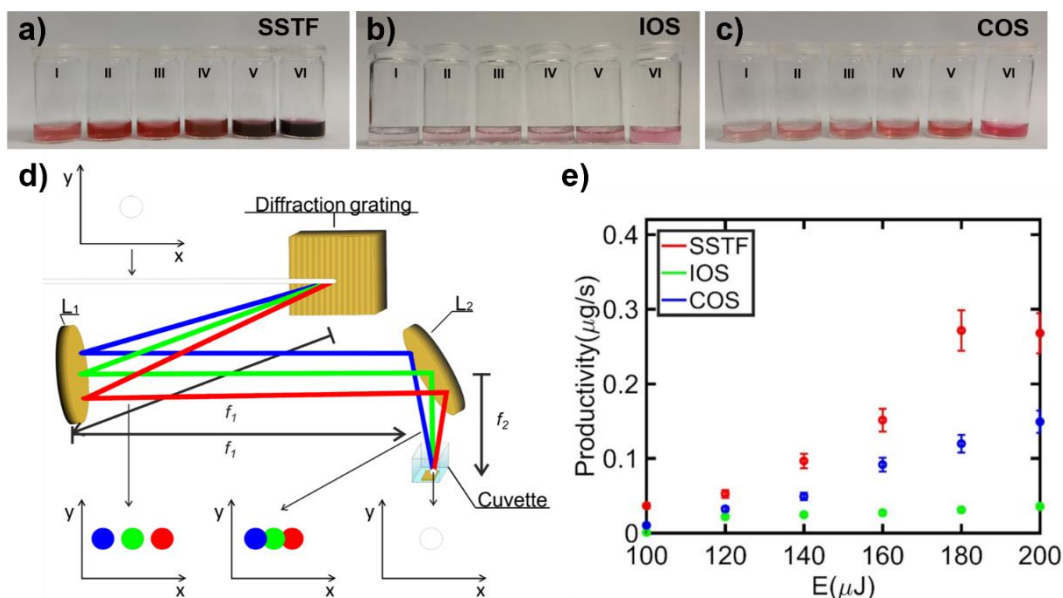

**Figure 23.** Image of Au generated colloids as a function of the pulse energy I=100 µJ, II=120 µJ, III=140 µJ, IV=160 µJ, V=180 µJ, VI=200 µJ, for a) SSTF, b) IOS and c) COS systems. d) Implemented experimental setup of the SSTF technique for femtosecond laser ablation in liquid. e) Productivity values obtained from the a), b) and c) colloids proving the enhanced SSTF production. The figure is reprinted from [60], Copyright 2019, under Creative Commons Attribution. Retrieved from https://doi.org/10.1364/PRJ.7.001249.

The proposed SSTF Au NP production, Figure 23a, is compared against an analogous image system (IOS) without temporal focusing effect, Figure 23b, based on a 4f system that collimates the beam, followed by focusing with a spherical lens, and the conventional PLAL setup (COS) directly using the beam output and focusing it with a standard spherical lens, Figure 23c. The characterization of the energy losses through different water layers confirms a maximum energy loss of 5% for the SSTF setup, 40% for the IOS system and 70% for the COS, showing the large reduction of nonlinear effects when employing SSTF PLAL compared to the standard femtosecond PLAL (COS). The increase of the energy delivered to the target confirms that, as for ablation in air, femtosecond pulses represent an efficient pulse duration for ablation processes. The Au NP productivity is increased a factor ≈2 compared to the standard femtosecond PLAL system even when the experimental parameters as fluence are favourable for the conventional femtosecond PLAL system. When compared with a system with the same parameters (IOS) the NP productivity increase factor is enhanced to ≈10 [60].

- Confines femtosecond pulse intensity to focal volume, minimizing nonlinear losses in the liquid where the pulses have a longer duration.
- Improves energy delivery leading to a productivity boost of a factor 2 compared to conventional fs-PLAL and a factor 10 compared to an analogous optical system.
- Requires diffractive grating and precise alignment, increasing system complexity.

### 5.2. Multi-beam PLAL

After exploring the spatial and temporal beam influence on PLAL, an unexplored limitation is the employment of a single beam. Generally, using a single beam in laser applications such as material processing and NP synthesis limits the processing speed and NP yield. Yet, alternatives to the single beam had not been explored for PLAL. The multi-beam (MB) irradiation for PLAL has been proposed employing a diffractive optical element (DOE) to split the laser beam into several sub-beams [210] to reduce energy losses associated for example with a programmable spatial light modulator [211]. Modern DOEs are flexible and reliable devices that enable the reshaping of a laser beam to almost any desirable distribution, e.g., creating a line or a 2D pattern of sub-beams, while maintaining the parameters of the light source (beam size, divergence, and polarization). The MB technique is currently widely used in various laser applications such as material processing [212,213], optical sensing [214,215], and lithography [216]. Recently, the first experiments on multi-beam laser ablation in liquids (MB-PLAL) were performed to investigate the possibility of increasing the productivity of the synthesis of colloidal alloy NPs by employing MB-PLAL [55,56]. Different DOEs with a splitting factor from 2 to 11 and a standard galvanometric scanner were used. The increased efficiency of MB-PLAL for NP production was demonstrated, Figure 24. With 11 beams, the obtained productivity was 3-4 times higher than that with the standard single-beam PLAL and, for FeNi NPs, it reached 1.6 g/h which is comparable in terms of production per watt with the record values for PLAL-produced NPs obtained with a fast polygon scanning system, which is expensive and not widely available [53]. The size distributions of the NPs obtained with MB and single-beam PLAL setups are found to be identical [55,56].

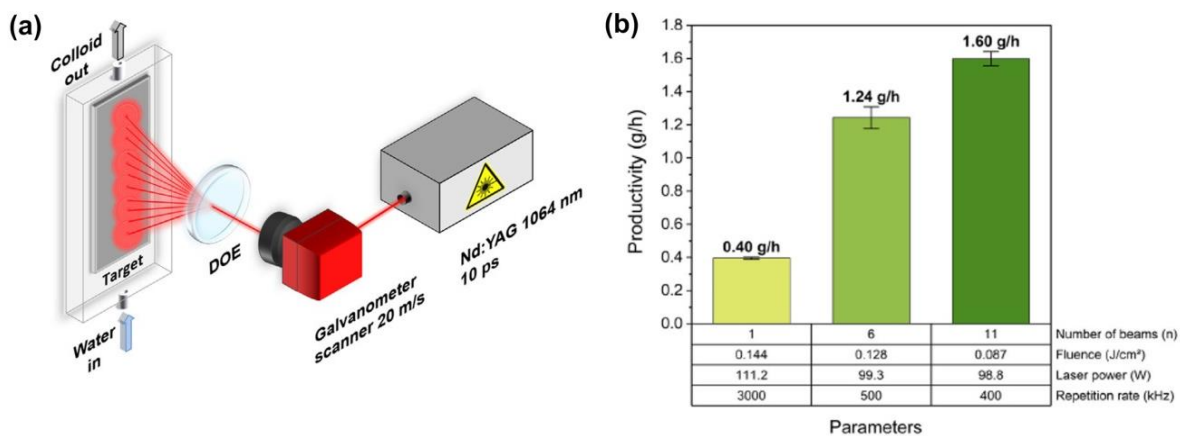

**Figure 24.** (a) Schematic illustration of the MB-PLAL process. (b) Productivity of FeNi NPs in water with a different number of laser beams. The figure is reprinted from [55], Copyright 2024, under Creative Commons Attribution. Retrieved from https://doi.org/10.1002/adpr.202300290.

It is important to underline that the increase in the NP productivity by MB-PLAL is not due to simply a larger number of laser beams but due to the possibility of bypassing the CB, the main limiting factor under single-beam PLAL [53]. To maintain the optimum fluence when the laser beam is split, the repetition rate is reduced to the same factor as the number of beams generated. The fact that the pulse repetition rate can be reduced proportionally to the DOE splitting factor with keeping the same number of pulses per time helps to bypass the bubble both temporally (the inter-pulse time can be shorter than the bubble lifetime) and spatially (the inter-pulse distance at a given scanning speed can be smaller than the bubble lateral size). Gatsa et al. [56] investigated in detail the effects of these two factors separately (keeping all the rest of PLAL parameters fixed) and demonstrated that, at certain conditions (e.g., very short inter-pulse distances), the gain in productivity with MB-PLAL can be even higher than the DOE beam splitting factor (examples of the dependencies for high entropy alloy NPs are shown in Figure 25). The obtained results suggest that the PLAL productivity of NPs at the level of several g/h can be routinely achieved by the multi-beam approach using modern compact kW-class lasers and simple inexpensive scanning systems.

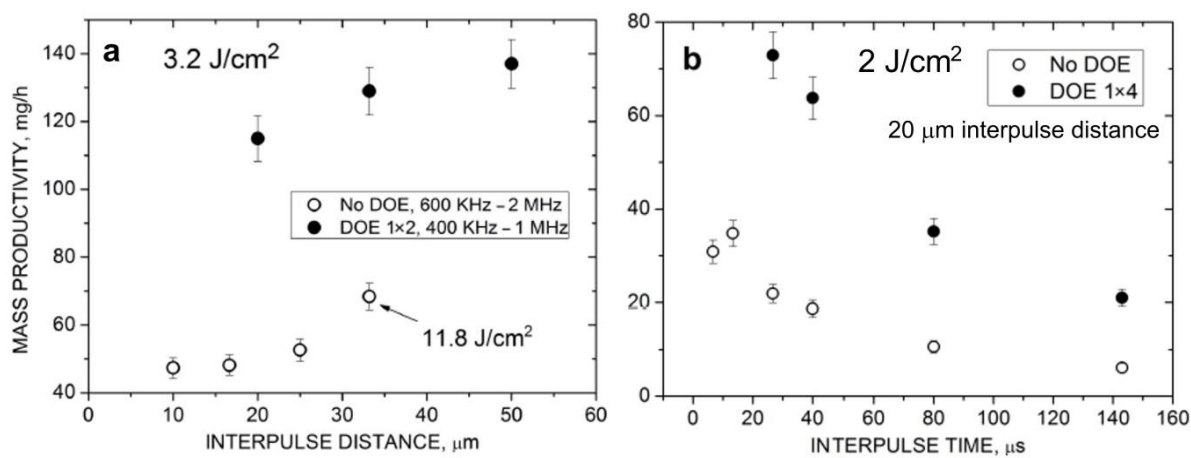

**Figure 25.** Productivity of CrFeCoNiMn high entropy alloy NPs by ps PLAL in water with and without DOEs as a function of (a) inter-pulse distance and (b) inter-pulse time. The figure is reprinted from [56], Copyright 2024, under Creative Commons Attribution. Retrieved from https://doi.org/10.3390/nano14040365.

- Splitting beam into 2–11 sub-beams increase productivity up to a factor of 4 for the tested materials and systems.
- Maintains identical NP size distribution compared to single-beam.
- Reduces the requirement of faster scanning systems and higher repetition rate lasers for PLAL gram per hour scale production.
- Requires a DOE and synchronization of the repetition rate with the beam-splitting factor.
- System complexity not significantly increased, and scalable with compact kW-class lasers.

SSTF represents a major step forward for femtosecond PLAL. By spatially chirping the beam so that the shortest pulse duration is recovered only at the focal spot, SSTF suppresses nonlinear effects such as filamentation and optical breakdown. Experimental data confirm that energy losses in water drop dramatically with SSTF (≈5%) compared to IOS (≈40%) and conventional PLAL (≈70%), enabling more efficient energy delivery. As a result, Au NP productivity increases a factor 2 relative to standard femtosecond PLAL and 10 times compared to IOS systems under the same parameters [60]. MB-PLAL further enhances scalability by splitting a single beam into multiple sub-beams using DOEs. Productivity scales nearly linearly with the number of beams, reaching 1.6 g·h⁻¹ for FeNi NPs with 11 beams with simple and accessible equipment. Critically, MB-PLAL bypasses cavitation bubble limitations both temporally and spatially, allowing repetition rate reduction while maintaining the same total number of pulses. Under short inter-pulse distances, productivity gains can even exceed the DOE splitting factor, indicating strong potential for routine NP production in the multi-gram-per-hour range using compact kW-class lasers [55].

Together, SSTF and MB-PLAL demonstrate that combining spatial and temporal control is a powerful route toward highly productive, energy-efficient, and scalable PLAL systems without compromising NP size distribution.

## Future Perspectives

The production of NPs via PLAL needs careful consideration of multiple parameters, including laser fluence, irradiance, focal length, beam shape, temporal pulse range and repetition rate, among others, to thoroughly evaluate and optimize productivity. A key advantage of PLAL over chemical synthesis lies in its simplicity and versatility, requiring only basic components such as a laser, scanner, chamber, liquid, and target. However, unresolved issues still need to be addressed to unlock the full potential of PLAL and enable further advancements in its scalability and efficiency for broader industrial use. The current review covers beam shaping strategies towards the overcoming of the current PLAL limitations. Further approaches related to beam shaping and beam control strategies are yet to be explored, representing an opened research field towards the industrialization of PLAL and its broadening of applications:

**Raster scanning speed-up:** Raster and galvanometer laser scans are growing at a breakneck pace, driven by emerging trends such as machine learning and digitalization of experiments [53,54]. As it is possible to obtain better control and energy distribution on the sample, the production rates of NPs have considerably increased. These technologies aim to improve motion control, refine trajectory planning, and facilitate predictive maintenance, enhancing both accuracy and reliability. The production of pulsed laser nanomaterials in liquid media could achieve significant qualitative and quantitative advancements by correcting optical aberrations introduced by optical elements. This would enable the development of advanced focusing techniques, such as processing with focal structures generated by cylindrical lenses or optimizing lens types for immersion in liquids. Integrating high-speed processing systems with application-specific, customized optical configurations presents a promising approach to addressing persistent challenges that currently limit the efficiency of PLAL.

**Smart beam optical delivery systems:** By integrating advanced beam delivery systems, the aim is to reduce the stiffness of the system by mitigating non-linear effects, using SSTF, as well as to spatially structure the beam both statically, using diffractive optical elements or free-form optics, and dynamically, using Spatial Light Modulators or Digital Micromirror Devices. Normally, the high-power laser used in PLAL exceeds, by several orders of magnitude, the optimum energy required for laser ablation. Therefore, parallel processing [217], different spatial profiles [218] or systems for focusing through turbid media [219] present an effective alternative to enhanced NP yield. Furthermore, the application of metamaterials in NP synthesis presents significant potential. Metamaterials can be engineered to precisely control light-matter interactions, optimizing laser energy absorption in the target material and enhancing ablation efficiency. Moreover, metamaterials exhibiting negative refractive indices or plasmonic resonances can be strategically designed to localize and intensify energy deposition in the target, enabling more controlled and efficient ablation processes. The PLAL scientific community must keep abreast of technologies developed in other fields that may be adaptable for PLAL applications.

**Boost efficiency across different pulse duration ranges:** Knowing and using the appropriate temporal range for each application will not only allow the generation and modification of nanomaterials with specific properties but will also reduce or even eliminate the post-processing steps common in industrial applications, reducing costs and production time. The duration of the laser pulse plays a crucial role in the synthesis of nanocomposites, affecting nearly every stage of their formation, including morphology, composition, atomic redistribution, structural shape, oxidation processes, vacancy generation, and crystallization dynamics [220–225]. To precisely control PLAL nanomaterials, enhancing productivity across various pulse width regimes is crucial, rather than limiting high production to picosecond pulses alone. Additionally, it is worth noting that the potential of extremely short pulse durations, such as those in the attosecond range, remains unexplored for NP generation in PLAL.

**Tailoring strategies in temporal regime:** Tailoring the temporal pulse envelope of ultrashort pulses could be introduced to the PLAL method. While existing techniques, such as double-pulse laser irradiation [204–206] or burst pulse irradiation, still requires further investigation to optimize productivity. The door is now open for further research in more complex areas such as ablation with shorter pulses in the attosecond regime. Furthermore, the ability to shape the temporal profile of laser pulses offers a means to fine-tune energy delivery rates to align with specific material reactions, highlighting the need for continued development of these temporal beam manipulation techniques.

**Automatization of NP synthesis:** Automated processes significantly outperform humans in speed, and feedback systems enable real-time adjustments during fabrication. Process automation in tasks like liquid refilling, beam focalization, and component cleaning reduces variations caused by human error. This also lessens the labour-intensive nature of PLAL, saving time that would otherwise be lost to production halts and reducing downtime. While some studies have explored off-site fabrication monitoring and management [15], achieving broader commercialization of PLAL will require higher levels of automation. In this way, analyzing datasets on target material properties, laser parameters, and liquid media could be automatized by artificial intelligence algorithms [226]. The development of dedicated digital tools for augmenting the research outcome and reproducibility can determine the best conditions for NP synthesis. Additionally, this enhanced understanding of optimal PLAL conditions can accelerate production scaling, reduce trial-and-error experimentation and enable large-scale output to meet market demands.

**Fluid mechanics and ablation chamber designs:** While various fluid dynamics strategies have been developed to enhance production rates, such as modifications to the ablation chamber, there is still potential for improvement. In this regard, improved fluid mechanics simulations can optimize not only the chamber design but also the management of turbulence, bubble removal, reduction of NP accumulation areas, and the selection of the liquid for appropriate applications. On the other hand, the use of 3D printing offers an innovative alternative to manufacturing ablation chambers in PLAL, allowing customized designs that optimize fluid flow, reduce material redeposition and improve NPs stability. It also facilitates the integration of channels, filters and surfaces with specific properties to modulate process dynamics. The use of advanced laser-compatible materials and the possibility of rapid prototyping speed up system experimentation and optimization, even going as far as to automatize the procedure of designing a chamber. Overall, this technology improves the efficiency, reproducibility and control of the nanoparticle synthesis process.

The growing global population and its escalating demands necessitate the development of more efficient methods for nanomaterial synthesis. PLAL has emerged as a versatile and promising technique, with extensive applications and significant support from the scientific community focused on advancing eco-friendly fabrication technologies. Enhancing the productivity of PLAL is a critical objective, as it could significantly lower the cost and increase the accessibility of sustainable nanomaterial production for diverse stakeholders, including researchers, industries, and consumers. Furthermore, improving PLAL productivity will expand its applicability, unlocking new opportunities for innovation.

## Acknowledgements

The authors are very grateful to the "Serveis Centrals d'Instrumentació Científica" (SCIC) of the University Jaume I. The graphical abstract is reprinted from [5], Copyright 2023, under Creative Commons Attribution. Retrieved from https://doi.org/10.1039/D3CP01214J.

# Fundings

This work has been funded by Ministerio de Ciencia e Innovación with grant PRE2020-094889 funded by MCIN/AEI/ 10.13039/501100011033, "ESF Investing in your future". The project PID2022-142907OB-I00 funded by MCIN/AEI/10.13039/501100011033 and "ERDF: A Way of Making Europe" and research project PID2023-153196OB-I00 funded by MICIU/AEI/10.13039/501100011033, by FEDER, and UE. N.M.B. and A.V.B acknowledge funding from the European Regional Development Fund and the State Budget of the Czech Republic (project SenDiSo: CZ.02.01.01/00/22_008/0004596). A.V.B acknowledges support from the Czech Science Foundation (GAČR), project 22-38449L. C.D-B. acknowledges support from Generalitat Valenciana (CIDEIG/2023/08) and University Jaume I (UJI-2024-21). S. M-P., J.L. and G.M-V. thank Generalitat Valenciana for funding project CIPROM/2023/44.